\documentstyle[12pt]{article}
\begin{document}
\makeatletter
%
%
%
%
\newdimen\normalarrayskip              
\newdimen\minarrayskip                 
\normalarrayskip\baselineskip
\minarrayskip\jot
\newif\ifold             \oldfalse
\newif\ifdisplayarray    \displayarraytrue
\newif\ifbigarray        \bigarraytrue
\def\arraymode{\ifold\relax\else\ifdisplayarray\displaystyle\else\relax\fi\fi} 
\def\eqnumphantom{\phantom{(\theequation)}}     
\def\@arrayskip{\ifold\baselineskip\z@\lineskip\z@\else\ifbigarray
     \baselineskip\normalarrayskip\lineskip\minarrayskip
     \else
     \baselineskip\z@\lineskip\z@\fi\fi}
\def\@arrayclassz{\ifcase \@lastchclass \@acolampacol \or
\@ampacol \or \or \or \@addamp \or
   \@acolampacol \or \@firstampfalse \@acol \fi
\edef\@preamble{\@preamble
  \ifcase \@chnum
     \hfil$\relax\arraymode\@sharp$\hfil
     \or $\relax\arraymode\@sharp$\hfil
     \or \hfil$\relax\arraymode\@sharp$\fi}}
\def\@array[#1]#2{\setbox\@arstrutbox=\hbox{\vrule
     height\arraystretch \ht\strutbox
     depth\arraystretch \dp\strutbox
     width\z@}\@mkpream{#2}\edef\@preamble{\halign \noexpand\@halignto
\bgroup \tabskip\z@ \@arstrut \@preamble \tabskip\z@ \cr}%
\let\@startpbox\@@startpbox \let\@endpbox\@@endpbox
  \if #1t\vtop \else \if#1b\vbox \else \vcenter \fi\fi
  \bgroup \let\par\relax
  \let\@sharp##\let\protect\relax
  \@arrayskip\@preamble}
%
%
%
\def\eqnarray{\stepcounter{equation}%
              \let\@currentlabel=\theequation
              \global\@eqnswtrue
              \global\@eqcnt\z@
              \tabskip\@centering
              \let\\=\@eqncr
              $$%
 \halign to \displaywidth\bgroup
    \eqnumphantom\@eqnsel\hskip\@centering
    $\displaystyle \tabskip\z@ {##}$%
    &\global\@eqcnt\@ne \hskip 2\arraycolsep
         \hfil$\arraymode{##}$\hfil
    &\global\@eqcnt\tw@ \hskip 2\arraycolsep
         $\displaystyle\tabskip\z@{##}$\hfil
         \tabskip\@centering
    &{##}\tabskip\z@\cr}
%
%
%
\newenvironment{marray}{\begin{equation}\begin{array}}%
{\end{array}\end{equation}}
%
%
\newenvironment{carray}{\begin{equation}\begin{array}{rcl}}%
{\end{array}\end{equation}}
\def\be{\@ifnextchar[{\def\ee{\end{equation}}\begin{equation}\l@b}%
{\def\ee{$$}$$}}
\def\l@b[#1]{\label{#1}}
\def\ba{\@ifnextchar[{\def\ee{\end{carray}}\begin{carray}\l@b}%
{\def\ee{\end{array}$$}$$\begin{array}{rcl}}}
\def\barray#1{\@ifnextchar[{\def\ee{\end{marray}}\begin{marray}{#1}\l@b}%
{\def\ee{\end{array}$$}$$\begin{array}{#1}}}
%
%
%
\def\herring{\@ifnextchar[{\@herring}{\@herring[\vcenter]}}
\def\@herring[#1]#2{\begingroup
\def\*{\\ \>}
\topsep0pt
\partopsep0pt
\def\tabbing{\lineskip\jot \lineskiplimit\jot
     \let\>\@rtab\let\<\@ltab\let\=\@settab
     \let\+\@tabplus\let\-\@tabminus\let\`\@tabrj\let\'\@tablab
     \let\\=\@tabcr
     \global\@hightab\@firsttab
     \global\@nxttabmar\@firsttab
     \dimen\@firsttab\@totalleftmargin
     \global\@tabpush0 \global\@rjfieldfalse
     \trivlist \item[]\if@minipage\else\vskip\parskip\fi
     \setbox\@tabfbox\hbox{\rlap{\indent\hskip\@totalleftmargin
       \the\everypar}}\def\@itemfudge{\box\@tabfbox}\@startline\ignorespaces}
\def\@startfield{\global\setbox\@curfield\hbox
                    \bgroup$\displaystyle}%
\def\@stopfield{$\egroup}%
#1{\begin{tabbing}#2\end{tabbing}}\endgroup}
%
%
%
\def\boldbox#1{{\mathsurround0pt\mathchoice{\hbox{\boldmath $#1$}}%
{\hbox{\boldmath $#1$}}{\hbox{\boldmath $\scriptstyle#1$}}
{\hbox{\boldmath $\scriptscriptstyle#1$}}}}
%
%
%
\def\textbox#1{{\mathchoice{\mbox{#1}}{\mbox{#1}}%
{\mbox{{\scriptsize#1}}}{\mbox{{\tiny#1}}}}}
%
%
%
%
\def\sect#1{\ref{#1}}
\def\eq#1{(\ref{#1})}
\def\theequation{\thesection.\arabic{equation}}
\@addtoreset{equation}{section}%
\def\@cite#1#2{\hbox{ [#1\if@tempswa ,#2\fi]}}
\def\@citex[#1]#2{\if@filesw\immediate\write\@auxout{\string\citation{#2}}\fi
  \def\@citea{}\@cite{\@for\@citeb:=#2\do
    {\@citea\def\@citea{,\penalty\@m}\@ifundefined  
       {b@\@citeb}{{\bf ?}\@warning
       {Citation `\@citeb' on page \thepage \space undefined}}%
\hbox{\csname b@\@citeb\endcsname}}}{#1}}
%
\def\@sect#1#2#3#4#5#6[#7]#8{\ifnum #2>\c@secnumdepth
     \def\@svsec{}\else
     \refstepcounter{#1}\edef\@svsec{\csname the#1\endcsname.%
     \hskip 0.8em }\fi
     \@tempskipa #5\relax
      \ifdim \@tempskipa>\z@
        \begingroup #6\relax
          \@hangfrom{\hskip #3\relax\@svsec}{\interlinepenalty \@M #8\par}%
        \endgroup
       \csname #1mark\endcsname{#7}\addcontentsline
         {toc}{#1}{\ifnum #2>\c@secnumdepth \else
                      \protect\numberline{\csname the#1\endcsname}\fi
                    #7}\else
        \def\@svsechd{#6\hskip #3\@svsec #8\csname #1mark\endcsname
                      {#7}\addcontentsline
                           {toc}{#1}{\ifnum #2>\c@secnumdepth \else
                             \protect\numberline{\csname the#1\endcsname}\fi
                       #7}}\fi
     \@xsect{#5}}
\newenvironment{appendices}{\begingroup
\setcounter{subsection}{0}
\def\thesubsection{\Alph{subsection}}%
\def\theequation{\thesubsection.\arabic{equation}}%
\@addtoreset{equation}{subsection}%
\def\appendix##1{\subsection{##1}}%
}{\endgroup}
\makeatother
%
\def\note#1{\typeout{#1}}
\def\BUG#1{\vrule width 2pt height 8pt depth 2pt\relax
\typeout{BUG? page= \thepage: #1}}
\def\dubious{\typeout{DUBIOUS: page= \thepage}}
\newcommand {\ignore}[1]{}
\newcommand{\nota}[1]{\makebox[0pt]{\,\,\,\,\,/}#1}
\newcommand{\notp}[1]{\makebox[0pt]{\,\,\,\,/}#1}
\newcommand{\braket}[1]{\mbox{$<$}#1\mbox{$>$}}
\newcommand{\Frac}[2]{\frac{\displaystyle #1}{\displaystyle #2}}
\renewcommand{\arraystretch}{1.5}
\newcommand{\noi}{\noindent}
\newcommand{\bc}{\begin{center}}
\newcommand{\ec}{\end{center}}
\newcommand{\epm}{e^+e^-}
\def\ifmath#1{\relax\ifmmode #1\else $#1$\fi}
\def\half{\ifmath{{\textstyle{1 \over 2}}}}
\def\quarter{\ifmath{{\textstyle{1 \over 4}}}}
\def\3quarter{{\textstyle{3 \over 4}}}
\def\third{\ifmath{{\textstyle{1 \over 3}}}}
\def\twothirds{{\textstyle{2 \over 3}}}
\def\fourth{\ifmath{{\textstyle{1\over 4}}}}
\def\sqrthalf{\ifmath{{\textstyle{1\over\sqrt2}}}}
\def\halfsqrthalf{\ifmath{{\textstyle{1\over2\sqrt2}}}}
\def\cl{\centerline}
\def\vs{\vskip}
\def\hs{\hskip}
\def\ra{\rightarrow}
\def\Ra{\Rightarrow}
\def\us{\undertext}
\overfullrule 0pt
\def\lf{\leaders\hbox to 1em{\hss.\hss}\hfill}
\def\21{$SU(2) \ot U(1)$}
\def\321{$SU(3) \ot SU(2) \ot U(1)$}
\def\ne{\hbox{$\nu_e$ }}
\def\nm{\hbox{$\nu_\mu$ }}
\def\nt{\hbox{$\nu_\tau$ }}
\def\ns{\hbox{$\nu_{s}$ }}
\def\nx{\hbox{$\nu_x$ }}
\def\Nt{\hbox{$N_\tau$ }}
\def\nr{\hbox{$\nu_R$ }}
\def\O{\hbox{$\cal O$ }}
\def\L{\hbox{$\cal L$ }}
\def\mne{\hbox{$m_{\nu_e}$ }}
\def\mnm{\hbox{$m_{\nu_\mu}$ }}
\def\mnt{\hbox{$m_{\nu_\tau}$ }}
\def\mq{\hbox{$m_{q}$}}
\def\ml{\hbox{$m_{l}$}}
\def\mup{\hbox{$m_{u}$}}
\def\md{\hbox{$m_{d}$}}
%
\def\ie{\hbox{\it i.e., }}        \def\etc{\hbox{\it etc. }}
\def\eg{\hbox{\it e.g., }}        \def\cf{\hbox{\it cf.}}
\def\etal{\hbox{\it et al., }}
\def\neus{\hbox{neutrinos }}
\def\gau{\hbox{gauge }}
\def\neu{\hbox{neutrino }}
\def\c{\mathop{\cos \theta }}
\def\s{\mathop{\sin \theta }}
\def\tr{\mathop{\rm tr}}
\def\Tr{\mathop{\rm Tr}}
\def\Im{\mathop{\rm Im}}
\def\Re{\mathop{\rm Re}}
\def\bR{\mathop{\bf R}}
\def\bC{\mathop{\bf C}}
\def\eq#1{{eq. (\ref{#1})}}
\def\Eq#1{{Eq. (\ref{#1})}}
\def\Eqs#1#2{{Eqs. (\ref{#1}) and (\ref{#2})}}
\def\Eqs#1#2#3{{Eqs. (\ref{#1}), (\ref{#2}) and (\ref{#3})}}
\def\Eqs#1#2#3#4{{Eqs. (\ref{#1}), (\ref{#2}), (\ref{#3}) and (\ref{#4})}}
\def\eqs#1#2{{eqs. (\ref{#1}) and (\ref{#2})}}
\def\eqs#1#2#3{{eqs. (\ref{#1}), (\ref{#2}) and (\ref{#3})}}
\def\eqs#1#2#3#4{{eqs. (\ref{#1}), (\ref{#2}), (\ref{#3}) and (\ref{#4})}}
\def\fig#1{{Fig. (\ref{#1})}}
\def\partder#1#2{{\partial #1\over\partial #2}}
\def\secder#1#2#3{{\partial^2 #1\over\partial #2 \partial #3}}
\def\bra#1{\left\langle #1\right|}
\def\ket#1{\left| #1\right\rangle}
\def\VEV#1{\left\langle #1\right\rangle}
\let\vev\VEV
\def\gdot#1{\rlap{$#1$}/}
\def\abs#1{\left| #1\right|}
\def\pri#1{#1^\prime}
\def\ltap{\raisebox{-.4ex}{\rlap{$\sim$}} \raisebox{.4ex}{$<$}}
\def\gtap{\raisebox{-.4ex}{\rlap{$\sim$}} \raisebox{.4ex}{$>$}}
\def\lsim{\raise0.3ex\hbox{$\;<$\kern-0.75em\raise-1.1ex\hbox{$\sim\;$}}}
\def\gsim{\raise0.3ex\hbox{$\;>$\kern-0.75em\raise-1.1ex\hbox{$\sim\;$}}}
\def\half{{1\over 2}}
\def\beq{\begin{equation}}
\def\eeq{\end{equation}}
\def\bef{\begin{figure}}
\def\eef{\end{figure}}
\def\bet{\begin{table}}
\def\eet{\end{table}}
\def\bea{\begin{eqnarray}}
\def\ba{\begin{array}}
\def\ea{\end{array}}
\def\bi{\begin{itemize}}
\def\ei{\end{itemize}}
\def\ben{\begin{enumerate}}
\def\een{\end{enumerate}}
\def\ra{\rightarrow}
\def\ot{\otimes}
%
\def\com#1#2{
        \left[#1, #2\right]}
\def\eea{\end{eqnarray}}
\def\bentarrow{\:\raisebox{1.3ex}{\rlap{$\vert$}}\!\rightarrow}
\def\longbent{\:\raisebox{3.5ex}{\rlap{$\vert$}}\raisebox{1.3ex}%
        {\rlap{$\vert$}}\!\rightarrow}
\def\onedk#1#2{
        \begin{equation}
        \begin{array}{l}
         #1 \\
         \bentarrow #2
        \end{array}
        \end{equation}
                }
\def\dk#1#2#3{
        \begin{equation}
        \begin{array}{r c l}
        #1 & \rightarrow & #2 \\
         & & \bentarrow #3
        \end{array}
        \end{equation}
                }
\def\dkp#1#2#3#4{
        \begin{equation}
        \begin{array}{r c l}
        #1 & \rightarrow & #2#3 \\
         & & \phantom{\; #2}\bentarrow #4
        \end{array}
        \end{equation}
                }
\def\bothdk#1#2#3#4#5{
        \begin{equation}
        \begin{array}{r c l}
        #1 & \rightarrow & #2#3 \\
         & & \:\raisebox{1.3ex}{\rlap{$\vert$}}\raisebox{-0.5ex}{$\vert$}%
        \phantom{#2}\!\bentarrow #4 \\
         & & \bentarrow #5
        \end{array}
        \end{equation}
                }
%
%
%
\def\ap#1#2#3{           {\it Ann. Phys. (NY) }{\bf #1} (19#2) #3}
\def\arnps#1#2#3{        {\it Ann. Rev. Nucl. Part. Sci. }{\bf #1} (19#2) #3}
\def\cnpp#1#2#3{        {\it Comm. Nucl. Part. Phys. }{\bf #1} (19#2) #3}
\def\apj#1#2#3{          {\it Astrophys. J. }{\bf #1} (19#2) #3}
\def\app#1#2#3{          {\it Astropart. Phys. }{\bf #1} (19#2) #3}
\def\asr#1#2#3{          {\it Astrophys. Space Rev. }{\bf #1} (19#2) #3}
\def\ass#1#2#3{          {\it Astrophys. Space Sci. }{\bf #1} (19#2) #3}
\def\aa#1#2#3{          {\it Astron. \& Astrophys.  }{\bf #1} (19#2) #3}
\def\apjl#1#2#3{         {\it Astrophys. J. Lett. }{\bf #1} (19#2) #3}
\def\ap#1#2#3{         {\it Astropart. Phys. }{\bf #1} (19#2) #3}
\def\ass#1#2#3{          {\it Astrophys. Space Sci. }{\bf #1} (19#2) #3}
\def\jel#1#2#3{         {\it Journal Europhys. Lett. }{\bf #1} (19#2) #3}
\def\ib#1#2#3{           {\it ibid. }{\bf #1} (19#2) #3}
\def\nat#1#2#3{          {\it Nature }{\bf #1} (19#2) #3}
\def\nps#1#2#3{        {\it Nucl. Phys. B (Proc. Suppl.) }{\bf #1} (19#2) #3} 
\def\np#1#2#3{           {\it Nucl. Phys. }{\bf #1} (19#2) #3}
\def\pl#1#2#3{           {\it Phys. Lett. }{\bf #1} (19#2) #3}
\def\pr#1#2#3{           {\it Phys. Rev. }{\bf #1} (19#2) #3}
\def\prep#1#2#3{         {\it Phys. Rep. }{\bf #1} (19#2) #3}
\def\prl#1#2#3{          {\it Phys. Rev. Lett. }{\bf #1} (19#2) #3}
\def\pw#1#2#3{          {\it Particle World }{\bf #1} (19#2) #3}
\def\ptp#1#2#3{          {\it Prog. Theor. Phys. }{\bf #1} (19#2) #3}
\def\jppnp#1#2#3{         {\it J. Prog. Part. Nucl. Phys. }{\bf #1} (19#2) #3}
\def\rpp#1#2#3{         {\it Rep. on Prog. in Phys. }{\bf #1} (19#2) #3}
\def\ptps#1#2#3{         {\it Prog. Theor. Phys. Suppl. }{\bf #1} (19#2) #3}
\def\rmp#1#2#3{          {\it Rev. Mod. Phys. }{\bf #1} (19#2) #3}
\def\zp#1#2#3{           {\it Zeit. fur Physik }{\bf #1} (19#2) #3}
\def\fp#1#2#3{           {\it Fortschr. Phys. }{\bf #1} (19#2) #3}
\def\Zp#1#2#3{           {\it Z. Physik }{\bf #1} (19#2) #3}
\def\Sci#1#2#3{          {\it Science }{\bf #1} (19#2) #3}
\def\n.c.#1#2#3{         {\it Nuovo Cim. }{\bf #1} (19#2) #3}
\def\r.n.c.#1#2#3{       {\it Riv. del Nuovo Cim. }{\bf #1} (19#2) #3}
\def\sjnp#1#2#3{         {\it Sov. J. Nucl. Phys. }{\bf #1} (19#2) #3}
\def\yf#1#2#3{           {\it Yad. Fiz. }{\bf #1} (19#2) #3}
\def\zetf#1#2#3{         {\it Z. Eksp. Teor. Fiz. }{\bf #1} (19#2) #3}
\def\zetfpr#1#2#3{    {\it Z. Eksp. Teor. Fiz. Pisma. Red. }{\bf #1} (19#2) #3}
\def\jetp#1#2#3{         {\it JETP }{\bf #1} (19#2) #3}
\def\mpl#1#2#3{          {\it Mod. Phys. Lett. }{\bf #1} (19#2) #3}
\def\ufn#1#2#3{          {\it Usp. Fiz. Naut. }{\bf #1} (19#2) #3}
\def\sp#1#2#3{           {\it Sov. Phys.-Usp.}{\bf #1} (19#2) #3}
\def\ppnp#1#2#3{           {\it Prog. Part. Nucl. Phys. }{\bf #1} (19#2) #3}
\def\cnpp#1#2#3{           {\it Comm. Nucl. Part. Phys. }{\bf #1} (19#2) #3}
\def\ijmp#1#2#3{           {\it Int. J. Mod. Phys. }{\bf #1} (19#2) #3}
\def\ic#1#2#3{           {\it Investigaci\'on y Ciencia }{\bf #1} (19#2) #3}
\def\tp{these proceedings}
\def\pc{private communication}
\def\opc{\hbox{{\sl op. cit.} }}
\def\ip{in preparation}
\topmargin -2cm
\textwidth 15.5cm
\textheight 25.5cm
\renewcommand{\thefootnote}{\fnsymbol{footnote}}
\def\e{\mbox{e}}
\def\sgn{{\rm sgn}}
\def\gsim{\;
\raise0.3ex\hbox{$>$\kern-0.75em\raise-1.1ex\hbox{$\sim$}}\;}
\def\lsim{\;
\raise0.3ex\hbox{$<$\kern-0.75em\raise-1.1ex\hbox{$\sim$}}\;}
\def\MeV{\rm MeV}
\def\eV{\rm eV}
\thispagestyle{empty}
\begin{titlepage}
\begin{center}
\hfill hep-ph/9602307\\
\hfill FTUV/95-47\\
\hfill IFIC/95-49\\
\vskip 0.3cm
\large
{\bf The effect of random matter density perturbations
on the MSW solution to the solar neutrino problem }
\end{center}
\normalsize
\vskip1cm
\begin{center}
{\bf H. Nunokawa}
\footnote{
E-mail: nunokawa@flamenco.ific.uv.es, nunokawa@titan.ific.uv.es},
{\bf A. Rossi}
\footnote{
E-mail: rossi@evalvx.ific.uv.es, rossi@ferrara.infn.it},
{\bf V. B. Semikoz}
\footnote{E-mail: semikoz@evalvx.ific.uv.es;\\
On leave from the {\em Institute of the
Terrestrial Magnetism, the Ionosphere and Radio Wave Propagation of the
Russian Academy of Sciences, IZMIRAN, Troitsk, Moscow region, 142092, Russia}.}
{\bf and J. W. F. Valle}
\footnote{E-mail: valle@flamenco.ific.uv.es}\\
\end{center}
\begin{center}
\baselineskip=13pt
{\it Instituto de F\'{\i}sica Corpuscular - C.S.I.C.\\
Departament de F\'{\i}sica Te\`orica, Universitat de Val\`encia\\}
\baselineskip=12pt
{\it 46100 Burjassot, Val\`encia, SPAIN         }\\
\vglue 0.8cm
\end{center}

\begin{abstract}

We consider the implications of solar matter density random noise
upon resonant neutrino conversion. The evolution equation describing 
MSW-like conversion is derived in the framework of the Schr\"odinger 
approach. We study quantitatively their effect upon both large and 
small mixing angle MSW solutions to the solar neutrino problem. 
This is carried out both for the active-active $\nu_e \ra \nu_{\mu,\tau}$ 
as well as active-sterile $\nu_e \ra \nu_s$ conversion channels. 
We find that the small mixing MSW solution is much more stable 
(especially in $\Delta m^2$) than the large mixing solution. 
The possible existence of solar matter density noise at the 
few percent level could be tested at future solar neutrino 
experiments, especially Borexino.

\end{abstract}

\vfill

\end{titlepage}
\renewcommand{\thefootnote}{\arabic{footnote}}
\setcounter{footnote}{0}
\newpage

\section{Introduction}

The long-standing deficit of solar neutrinos (the Solar Neutrino 
Problem (SNP)) has now been observed by all four operating experiments 
\cite{cl,ga,sa,k,val95}.  The main essence of the SNP is the strong 
deficit of the beryllium neutrinos \cite{CF}. On the other hand, 
the high energy boron neutrinos are moderately suppressed, while 
the low energy ones are almost undepleted.  This strongly suggests 
that any astrophysical solution fails \cite{CF,BFL} in reconciling 
the experimental data with the Standard Solar Model (SSM) predictions 
\cite{SSM,turck,CDF}.  

It is possible to ascribe the solar neutrino deficit to the 
existence of two types of neutrino conversion mechanisms, both 
of which can deplete neutrinos of different energies differently, 
as required by the experimental data.
The long wavelength vacuum oscillations provide a good fit to 
the most recent results for $\Delta m^2\simeq 10^{-10}$eV$^2$ and 
large neutrino mixing $\sin^2 2 \theta \simeq 1$ \cite{KP,BR,Cala}.
The other scenario is the resonant neutrino conversion due to 
interactions with constituents of the solar material  
(the Mikheyev-Smirnov-Wolfenstein  (MSW) effect) \cite{MSW}.
This provides an extremely good data fit in the small mixing 
region with  $\Delta m^2 \simeq 10^{-5}$eV$^2$ and 
$\sin^2 2 \theta \simeq 10^{-3} \div 10^{-2}$ \cite{FIT,smirnov,Cala}. 
Both of these solutions have been studied against possible 
changes of the SSM input parameters \cite{smirnov,BR}. For example,
the study of the MSW effect has revealed its stability, especially 
in the $\Delta m^2$ parameter. 

In this paper we investigate the stability of the MSW solution with 
respect to the possible presence of random perturbations in the solar 
matter density, so far not included in the standard MSW picture.

In Ref.\cite{KS} the effect of periodic matter density perturbations 
added to an average density $\rho_0$, i.e.
\be[period]
\rho(r) = \rho_0  [1 + h \sin (\gamma r)]
\ee
upon resonant neutrino conversion was investigated. The major effects 
show up when the fixed frequency ($\gamma$) of the perturbation is close 
to the neutrino oscillation eigen-frequency, and for rather large amplitude 
values ($h \sim 0.1-0.2$), giving rise to the parametric effects \cite{KS}.  
Such effects can either enhance or suppress neutrino conversion in the Sun.
There are also a number of papers which address similar effects by different 
approaches \cite{AbadaPetcov,BalantekinLoreti}. 

Direct observations of solar surface motions, resulting from the 
superposition of several modes, may indicate a rich spectrum of frequencies. 
This would suggest the need to consider the effect of random or "white" 
noise matter density perturbations $\xi(r)$, characterised by an 
{\sl arbitrary} wave number $k$, 
\beq
\xi (r) = \int dk \xi(k)\sin kr  \:,
\eeq
rather than a periodic or regular perturbation. In such a case 
the spatial correlation function for a uniform medium
\be[corr]
\langle \xi (r_1)\xi (r_2)\rangle = \langle \xi^2
\rangle_{r_1 -r_2},
\ee
obeys $\langle \xi (k)\xi (k')\rangle = \langle \xi^2
\rangle_k \delta (k + k')$ as the averaging rule for the 
Fourier components, where the wave number $k$ is not fixed.
The effect of solar density as well as solar magnetic field 
fluctuations upon neutrino spin-flavour conversions has  
also been considered in Ref. \cite{BalantekinLoreti},
using somewhat different methods.

In this paper, after some discussion (Sec. 2) about the nature 
of the matter density fluctuations, we derive the most general 
neutrino evolution equation in random matter, starting from the 
standard Schr\"odinger equation (Sec. 3). This discussion is 
closer to the particle physics intuition than that of Ref. 
\cite{BalantekinLoreti}. Moreover, we consider both the 
active-active $\nu_e \rightarrow\nu_{\mu,\tau}$ as well as 
the active-sterile $\nu_e\rightarrow\nu_s$ neutrino conversion 
channels (here $\nu_s$ is a neutrino state with no standard model 
interaction). The latter is motivated by the fact that the existence
of a sterile neutrino seems to be the only way to simultaneously 
account for the solar and atmospheric neutrino deficits in the 
presence of neutrino dark matter \cite{DARK}.

After an analytical study of the neutrino conversion equations 
we have investigated the impact of matter density noise upon the 
MSW scenario in the context of the SNP (Sec. 4). Typically, we 
find that the presence of matter fluctuations weakens the MSW 
mechanism, thus reducing the resonant conversion probabilities
\cite{BalantekinLoreti}. We have carried out a fit of the latest 
solar \neu data for different values of the noise level, minimising 
the $\chi^2$ in the ($\Delta m^2, \sin^2 2 \theta$) plane. As in the 
noiseless MSW case, we find that the small mixing MSW solution 
provides a better fit to the data than the large mixing one, both 
for the case of active, as well as sterile neutrino conversions. 
We present several plots with the results of our fits in which 
the effect of the noise is studied in the idealised approximation 
where all neutrinos are produced at the solar centre. We conclude that 
the most relevant parameter region corresponding to adiabatic
conversion of $^7$Be neutrinos is relatively stable with respect 
to such density fluctuations, whereas there is a larger effect of the 
noise for the large mixing MSW solution. We show how the possible 
existence of solar matter density noise could be tested in the 
next generation of solar neutrino experiments, especially Borexino.
Finally, we comment on how possible solar model uncertainties could
affect our results.

\section{Matter Density Noise in the Sun }

Let us briefly discuss the expected size of fluctuations in the 
Sun and their correlation lengths.  For the sake of discussion, we 
can approximate (except in the very inner core) the average solar matter 
density, as given by the Standard Solar Model (SSM) \cite{SSM,turck,CDF}, 
by:
\be[profile]
\rho(r) \simeq \rho_0\exp (-\frac{r - r_0}{R_0})
\ee
where $R_0 \approx 0.1 R_s$ ($R_s$ is the solar radius),
$r_0 \simeq 0$, and $\rho_0 \approx 250 $g/cm$^3$. 
The SSM in itself cannot account for the existence of 
density perturbations, since it is based on hydrostatic 
evolution equations. 

One may however speculate upon possible mechanisms that
could induce such density inhomogeneities in the Sun. 
Unfortunately it is quite difficult to give reliable estimates of 
the density perturbations in deep layers of the Sun, since this would 
require the detection of $g$-modes, not yet possible \cite{gmode}.
Indeed, these modes can exist only in deep layers beneath the 
convective zone and thus they can reach the surface only after 
an exponential damping through the convective zone \cite{turck}.
Note also that it is extremely difficult to identify the $g$-modes 
in helioseismology observations, due to their tendency to be 
accumulated in the lower frequency part of the Fourier spectrum.
We may, however, give a simple estimate of the level of density 
perturbations $\delta \rho$ in the solar interior by combining 
the continuity equation up to the first order in $\delta \rho$ 
and the velocity perturbation $\delta \mbox{{\bf v}}$
\beq
\label{cont}
\frac{\partial \delta \rho}{\partial t} + 
\nabla\cdot \delta\mbox{{\bf v}}\rho = 0,
\eeq
with the $p$-mode observations of the IRIS network at Tenerife
\cite{turck}. These show that in the lower frequency part of the 
Fourier spectrum, the $p$-mode spectrum resembles that of noise,
namely $\sim 1/f$. For instance, using the measured power 
$\delta P = 10^3 \,\mbox{m}^2$s$^{-1}$ corresponding to the frequency
$f = 10^{-4}$ s$^{-1}$ (see Fig. 26 of Ref.\cite{turck}) we may
estimate from Eq. (\ref{cont}) the perturbation level $\xi$ defined as
\beq
\xi =\frac{\delta \rho}{\rho}\equiv 
\frac{\sqrt{\langle \delta \rho^2\rangle}}{\rho} .
\eeq
We obtain 
\beq
\xi^2 \simeq \frac{\delta v^2}{f^2L_0^2} = \frac{\delta P(f)}{fL_0^2} 
\sim \frac{10^3 \mbox{m}^2\mbox{s}^{-1}}{10^{-4}\mbox{s}^{-1}
 \times (10^6\mbox{m})^2} \sim 10^{-5}
\eeq
where we have taken as typical size of the spatial
inhomogeneity the value $L_0\sim 10^3$ km, the so-called
"granule"-size. Thus we see that values $\xi \sim 0.3$\% 
in the solar surface can not be excluded. In contrast, inside the 
solar core the estimate of the parameter $L_0$ becomes very rough.
In fact one expects that, due to buoyancy, the $g$-mode amplitudes 
beneath the convective zone can be larger than at the surface and, 
correspondingly, the inhomogeneity size $L_0$ smaller than at the edge 
of the Sun (see Fig. 13 (a) of Ref. \cite{turck}). As a result, for a 
fixed perturbation amplitude $\delta v^2 = const$ we can extrapolate 
the {\em continuous} power spectrum to low frequencies leading 
to a large density inhomogeneities since $\xi\sim f^{-1}L_0^{-1}$.

There is another way to estimate the level of density fluctuations 
using the density profile of Eq. (\ref{profile}). Indeed, in the 
hydro-dynamical approximation, density perturbations can be induced 
by corresponding temperature $T$ fluctuations due to convection of 
matter between layers with different local temperatures. For example, 
if we express the macroscopic matter density $\rho(r)$ through the 
Boltzmann distribution with the gravitational potential energy 
$U\!=\!m_p g(r) (r-r_0)$,  where $g(r) = G M(r)/r^2 $, $G$ being 
Newton's constant, $m_p$ the nucleon mass and $M(r)$  the mass 
contained in a sphere of radius $r$. The change $T \ra T+ \delta T$ 
leads to
\be
\rho(r; T + \delta T) = \rho_0 \exp 
\Bigl [-\frac{m_p g (r-r_0)}{T + \delta T}\Bigr ] = 
\rho (r;T) [ 1 + \xi ] , 
\ee
where $\xi \equiv \xi (r; T; \delta T)$. From this we have
\be[betalevel]
\xi 
= \frac{\delta \rho}{\rho} 
= \frac{m_p g (r-r_0)}{T} \frac{\delta T}{T} 
= \frac{(r - r_0)}{R_0} \frac{\delta T}{T},
\ee
where we have compared the relevant exponent with that in \eq{profile}.
One can argue that $\sqrt{\langle \delta T^2\rangle}/T \lsim 0.05$ 
is not in conflict neither with the SSM nor with present helioseismology 
observations \cite{turck,dal}. Thus from Eq. (\ref{betalevel}), taking 
$(r - r_0)/R_0 \sim 1$, we obtain a correspondingly comparable level 
of density fluctuations. 
Thus in what follows we assume the existence of such few percent 
level matter density fluctuations, up to 8\%
\footnote{Note that for an ideal plasma, like that in the Sun, 
the equilibrium plasma fluctuations are negligible \cite{Akhiezer}, 
$\xi = <\!\delta n^2_e\!>^{1/2}/n_e \sim (n_e r^3_{D})^{-1}\ll 1$, 
where $n_e$ is the electron density, $\delta n_e$ the corresponding 
fluctuation and $r_D= (T/4\pi e^2 n_e)^{1/2}$ the Debye radius. 
Since the number of particles inside the Debye radius is very 
large, $N_D= n_e r^3_D\gg 1$, with $r_D \sim 10^{-7}$ cm 
$ \ll l_{free}$ these fluctuations are irrelevant 
for our present discussion. }.

Now we generalise the above discussion to the case in which the 
perturbation $\delta \rho$ is of random nature. Following ref.
\cite{BalantekinLoreti} we assume that the random field 
$\delta \rho$ is a $\delta$-correlated Gaussian distribution.
For small inhomogeneities, the autocorrelation function 
$\langle \xi^2 \rangle$ can be taken as 
\be[correlator]
\langle \delta \rho(r_1)\delta \rho(r_2)\rangle = 2\rho^2\langle 
\xi^2\rangle L_0 \delta (r_1 - r_2)
\ee
whose correlation length $L_0$ obeys the following relation:
\be[size]
l_{{free}} \ll L_0 \ll \lambda_m
\ee
where $l_{\rm free}= (\sigma n_0)^{-1}$ is the mean free path 
of the electrons in the Sun. This lower bound is dictated by the 
hydro-dynamical approximation used later. For Coulomb interactions, 
the cross-section $\sigma$  is determined by the classical radius of 
electron $r_{0e} =e^2/m_ec^2\sim 2 \times 10^{-13}$cm, resulting in 
$l_{\rm free}\sim 10\,\mbox{cm}$ for a solar mean density 
$n_0 \sim 10^{24}$cm$^{-3}$ and $\sigma \sim 10^{-25}$cm$^2$. 
On the other hand, the upper bound expresses the fact that the scale of 
fluctuations should be much smaller than the characteristic neutrino 
matter oscillation length, $\lambda_m$, as indeed the $\delta$-correlation 
distribution in Eq. (\ref{correlator}) requires.

\section{Neutrino Conversion in Noisy Matter}

Let us consider a system of two neutrinos $\nu_{e}$ and 
$\nu_{x}$. In the case of active-active \neu conversion 
$x=\mu$ or $\tau$, while for the case of active-sterile \neu 
conversions $\nu_x=\nu_s$, with $\nu_s$ being the sterile state.

Neutrino propagation in the solar medium is affected by 
the coherent neutrino scattering off matter constituents 
which can be described in terms of the matter potential $V$. 
In the rest frame of the unpolarised matter, the potential 
is given, in the Standard Model, by 
\be[poten]
V = \frac{\sqrt{2}G_F}{m_p} \rho Y
\ee
where $G_F$ is the Fermi constant, $\rho$ is the matter density and 
$Y$ is a number  which depends on the \neu type and on the chemical 
content of the medium. More precisely, $Y= Y_e - \frac{1}{2}Y_n$ for 
the $\nu_e$ state, $Y= -\frac{1}{2}Y_n$ for \nm and \nt and $Y=0$ 
for the $\nu_s$ state, where $Y_{e,n}$ denotes the electron and 
neutron number per nucleon. The matter potential modifies the 
energy dispersion relations for neutrino states, leading to the 
phenomenon of resonant conversion (the MSW effect \cite{MSW}). 
Let us note that in this respect the potential $V$, i.e. 
the function $\rho$, previously described, represents an 
average macroscopic quantity.

Now we re-derive the evolution equation for the neutrino 
in the presence of matter density random perturbations,
which we regard as superimposed over the main average
matter density profile. It is clear from Eq. (\ref{correlator} )
that the random component of the potential can be written 
as $V(t) \xi $%
\footnote{The radial dependence of the solar matter density is 
understood as a time dependence since neutrinos are relativistic.}.

The evolution for the $\nu_e-\nu_y$ ($y=x$ or $y=s$)
system is governed by 
\barray{ll}[ev1]
i \frac{d}{dt}\left (\matrix{\nu_e \\ \nu_y}\right )= 
\left (\matrix{H_{e}  &  H_{e y} \\
            H_{ey} & H_{y}}\right )
 \left (\matrix{\nu_e \\ \nu_y}\right ), 
\ee
where the entries of the Hamiltonian matrix are given by
\footnote{In the Hamiltonian matrix, a term proportional 
to the identity has been removed.}
 \begin{eqnarray}
\label{matdef}
 & & H_e=  2 [A_{ey}(t) + \tilde{A}_{ey}(t)], ~~~~ H_y=0, \nonumber\\ 
  & & A_{ey}(t)  =  \frac{1}{2} [V_{ey}(t) 
 - \frac{\Delta m^2}{2E} \cos2 \theta], ~~~~~
\tilde{A}_{ey}(t) = \frac{1}{2} V_{ey}(t) \xi
\end{eqnarray}
Here $\theta$ is the neutrino mixing angle in vacuum,
$\Delta m^2$ is the \neu squared mass difference, and the 
matter potential for the active-active \neu conversion ($y=x$) reads 
\be[vex]
V_{ex}(t) = \frac{\sqrt{2} G_F}{m_p} \rho(t) (1-Y_n)
\ee
or alternatively in case of $\nu_s$
\be[vexs]
V_{es}(t) = \frac{\sqrt{2} G_F}{m_p} \rho(t) (1-\frac{3}{2}Y_n)
\ee
(the neutral matter relation $Y_e =1-Y_n$ has been used). 

The above system  can be rewritten in terms of the following 
equations:
\begin{eqnarray}
\label{sys}
\dot{P}(t)& =& 2H_{e{y}} I(t) \nonumber \\
\dot{R}(t)& = & - H_e(t)I(t) \nonumber \\
\dot{I}(t)& = & H_e(t)R(t) - H_{e{y}} (2P(t) - 1) 
\end{eqnarray} 
where $P = | \nu_e |^2$ is the $\nu_e$ survival probability,  
$R \equiv$ Re$(\nu_y^* \nu_e) $ and 
$I \equiv$ Im$(\nu_y^* \nu_e) $. 
The corresponding initial conditions are:
\be[ic]
P(t_0) = 1, \,\,\,\,\,\,\,\,I(t_0)=0, \,\,\,\,\,\,\,\,R(t_0) = 0 .
\ee
Defining
\begin{eqnarray}
R(t) \pm \mbox{i}I(t)& = &\mbox{e}^{\pm 
\mbox{i}\int_{t_0}^t H_e(t_1)dt_1} Z_{\pm}(t) \nonumber \\
Z_{\pm}(t) &= &\mp \mbox{i}\int^t_{t_0} H_{ey} (2P(t_1)-1)
\mbox{e}^{\mp\mbox{i}\int^{t_1}_{t_0}H_e(t_2)dt_2} dt_1 ,
\end{eqnarray}
we can express the auxiliary functions $R(t)$ and $I(t)$ as:
\begin{eqnarray}
\label{aux}
R(t)& = & \int^t_{t_0} H_{ey} (2P(t_1)-1) 
\sin(\int^t_{t_1}H_e(t_2)dt_2) dt_1 \\
I(t)& = & -\int^t_{t_0} H_{ey} (2P(t_1)-1) 
\cos(\int^t_{t_1}H_e(t_2)dt_2) dt_1 \label{aux1} .
\end{eqnarray}
After substituting the Eqs. (\ref{aux}) and  (\ref{aux1}) 
in the r.h.s  of (\ref{sys}), we can average over the 
random density distribution, taking into account that 
for the random component we  have:
\be[matcorrel]
\langle \tilde{A}_{ey}^{2n+1} \rangle  = 0, ~~~~~~
\langle \tilde{A}_{ey}(t)\tilde{A}_{ey}(t_{1}) \rangle = 
2 \kappa\delta (t - t_{1}) ,
\ee
where the quantity $\kappa$ is defined as
\be[den_noise]
\kappa(t)=  \langle \tilde{A}_{ey}^2(t)\rangle L_0 = \frac{1}{4} 
V^2_{ey}(t)
\langle \xi^2\rangle L_0 .
\ee
At this point all we need are the following averaged products 
\be[average]
\langle \tilde{A}_{ey}(t) R(t)\rangle = -\kappa(t) \langle I(t)\rangle,~~~~~~ 
\langle \tilde{A}_{ey}(t) I(t)\rangle = \kappa(t) \langle R(t)\rangle .
\ee
These are derived from Eqs. (\ref{aux}) and (\ref{aux1}) taking into account
also (\ref{vex}), and are correct up to $O(\kappa)$. In terms of the averaged 
quantities defined as $\langle P(t)\rangle = \cal {P}(t)$, 
$\langle R(t)\rangle = \cal {R}(t)$, $\langle I(t)\rangle = \cal {I}(t)$, 
we can write the noise-averaged variant of the set (\ref{sys}) as:
\begin{eqnarray}
\label{sys1}
\dot{\cal{P}}(t) &= &2 H_{ey} \cal{I}(t) \nonumber \\
\dot{\cal{R}}(t) & = & -2A_{ey}(t) \cal{I}(t) -2 \kappa(t)\cal{R}(t) 
 \nonumber \\
\dot{\cal{I}}(t) & = & 2A_{ey}(t) \cal{R}(t) -2 \kappa(t)\cal{I}(t) 
- H_{ey} (2 \cal{P}(t)-1) .
\end{eqnarray}
This system of equations explicitly exhibits the noise 
parameter $\kappa$
\footnote{These equations are equivalent to those obtained 
in Ref. \cite{BalantekinLoreti} in terms of the variables 
$x=2\cal{R}$,  $y=-2\cal{I}$ and $r=2\cal{P}-1$.}.
Eliminating $\cal{I}$ and $\cal{R}$ from  Eq. (\ref{sys1})
we can obtain the following third order differential 
equation for the averaged conversion probability $\cal{P}$:
\begin{eqnarray}
\label{m1}
&& A_{ey}(t)\frac{d^3\cal{P}(t)}{dt^3} + \Bigl [4\kappa(t)A_{ey}(t) - 
\dot{A}_{ey}(t)\Bigr ]
\frac{d^2\cal{P}}{dt^2}(t)
 + 
\Bigl [\omega_0^2(t)A_{ey}(t) + 2A_{ey}(t)\dot{\kappa}(t) \nonumber \\
 & &  - 2\kappa(t)\dot{A}_{ey}(t)\Bigr ]
\frac{d\cal{P}}{dt} 
 - 4 H^2_{e{y}}\cal{P}(t)\Bigl(\dot{A}_{ey}(t)-2\kappa(t)A_{ey}(t)\Bigr) 
= \nonumber \\
 & & - 2 H^2_{ey}\Bigl(\dot{A}_{ey}(t)-2\kappa(t)A_{ey}(t)\Bigr)  ,
\end{eqnarray}
where the frequency $\omega_0^2$ familiar from the MSW effect
is given as
\be[freq]
\omega_0^2(t) = 4 (A_{ey}^2(t) +  H^2_{e{y}}) ,
\ee
and  the initial conditions become:
\be[conditions1]
\cal{P}(t_0) = 1,  ~~~~~~
\dot{\cal{P}}(t_0) = 0,  ~~~~~~
\ddot{\cal{P}}(t_0) = 2 H^2_{ey} .
\ee
Let us notice that in the absence of noise ($\kappa=0$) the 
Eq. (\ref{m1}) 
reduces to the well known MSW  equation (cfr. with (Eq. (2.23) 
of the first paper in Ref. \cite{MSW}) with the change 
$\bar{H} = 2H_{ey}$, $H = 2A_{ey}$). 

In order to gain some more insight on the present picture
let us note that the MSW resonance condition,  
i.e. $A_{ey}(t) = V_{ey}(t) -\Delta m^2 \cos2\theta/2E = 0$,  
remains unchanged, 
due to the random nature of the matter perturbations. In other
words, the fact that the noise is a second order effect 
(see \eq{matcorrel}) means that it can only be seen in
the conversion probability. In order to ensure that the
correlation length $L_0$ is smaller than the neutrino 
wave length in the Sun, as required by the condition 
(\ref{size}), we choose to adjust $L_0$ as follows:
\be[L0]
L_0 = 0.1\times(\lambda_m)=0.1 \times \frac{2\pi}{\omega_0} .
\ee

In order to get a feeling for the importance of the noise term 
in the system (\ref{sys1}), note that the noise parameter $\kappa$ in 
\Eq{den_noise} is always smaller than $A_{ey}(t)$, for $\xi \lsim$
few \%, except at the resonance region. As a result, the density 
perturbation can have its maximal effect just at the resonance. 
However, this is not enough for the noise to give rise to sizeable 
effects. Since the noise term gives rise to a damping term in the system 
(\ref{sys1}), it follows that the corresponding noise length scale 
$1/\kappa$ be much smaller than the thickness of the resonance 
layer $\Delta r$. In other words, it is also necessary that the 
following {\it adiabaticity} condition 
\be[adiab]
\tilde{\alpha}_r= \Delta r\, \kappa_{res} > 1 ,
\ee
is satisfied. This condition is analogous to the standard 
MSW {\it adiabaticity} condition $\alpha_r > 1$ where 
$\alpha_r = \Delta r/(\lambda_m)_{res}$ is the standard 
adiabaticity parameter at resonance \cite{MSW}. 
One can show that the two adiabaticity parameters are related as 
\be[alfa] 
\tilde{\alpha}_r \approx \alpha_r\, \frac{\xi^2}{\tan^2 2\theta}, \,\,\,\,\,
\,\,\,\,\,\,\,\,
\alpha_r = \frac{\Delta m^2 \sin^2 2 \theta R_0}{4\pi E \cos 2\theta} .
\ee
For the range of parameters we are considering, $\xi \sim 10^{-2}$ 
and $\tan^2 2\theta\geq 10^{-3}-10^{-2}$, and due to the restriction 
in the r.h.s of 
(\ref{size}) one can estimate that $\tilde{\alpha}_r \leq \alpha_r$. 
Moreover, the relation  $\tilde{\alpha}_r \leq \alpha_r$ can be 
rewritten as $\kappa_{res} < \delta H_{r}$, where $\delta H_{r}$ 
is the level splitting between the energies of the neutrino mass 
eigenstates at resonance. This shows that the energy perturbation 
induced by the matter fluctuations is not enough to cause the 
level crossing (even at the resonance) \cite{KS}. 
In other words, it never
violates the MSW adiabaticity condition 
\footnote{This is opposite to the case of a local density jump as 
discussed by Krastev and Smirnov in the second paper in Ref.\cite{KS}, 
where larger values of $\delta \rho$ could break MSW adiabaticity.}.  

From \Eq{alfa} it follows that, in the adiabatic regime 
$\alpha_r >1$, the effect of the noise is larger the smaller
the mixing angle value. Furthermore, as already noted above,
\Eq{alfa} implies that the MSW non-adiabaticity $\alpha_r <1$ 
is always transmitted to $\tilde{\alpha}_r < 1$. As a result,
under our assumptions the fluctuations are expected to be 
ineffective in the non-adiabatic MSW regime. 
 
\section{MSW Effect in Noisy Solar Matter}

In this section we study the impact that random perturbations 
in the solar matter density can have upon the MSW solution to the
solar neutrino problem. We will consider both the active to active
and active to sterile neutrino conversion.
For definiteness we will take as our reference SSM the most recent 
Bahcall-Pinsonneault (BP95) model with helium and heavy element diffusion, 
as given in the last paper in Ref.\cite{SSM}. From there we will take 
both the electron (neutron) density as well as the neutrino energy spectra 
and detection cross sections. 
Using these as input, we have solved numerically the coupled 
differential equations  in (\ref{sys1}) for the $\nu_e$ survival 
probability
\footnote{For simplicity and CPU economy we have not included 
throughout our analysis the neutrino production distributions in the Sun.}.

In order to get some preliminary insight on the effect of
the density noise, in Fig.1 we plot  $\cal{P}$ as a function 
of $E/\Delta m^2$ for different values of the noise parameter $\xi$.
For comparison, the standard MSW case $\xi=0$ is also shown 
(lower solid curve). We take this case as the reference 
situation with which all others with non-vanishing $\xi$ are 
compared. 

One concludes that in both cases of small and large mixing 
(Fig. 1a and Fig. 1b, respectively), the effect of the matter
density noise is to weaken the MSW suppression in the adiabatic
regime (see dotted and dashed curves) with negligible effect in 
the non-adiabatic region, in complete agreement with the results 
of Ref.\cite{BalantekinLoreti}. The relative increase 
of the survival probability $\cal{P}$ is larger for the case 
of small mixing (Fig. 1a) as already guessed on the basis of 
Eq. (\ref{alfa}). One sees that the enhancement of the survival 
probability can easily reach 20\% for $\xi$ values as small as $4\%$. 
From these figures one can already infer that for the relevant 
$\Delta m^2 \sim 10^{-5}$ eV$^2$ the intermediate energy 
neutrinos (like $^7$Be neutrinos) are the ones most likely
to be affected by the matter noise. 

Note that the "white noise"-type density fluctuations we consider
here cannot lead to any parametric enhancement \cite{landau} 
of the survival probability  of the type discussed in Ref.\cite{KS} 
with a sinusoidal density perturbation. In contrast to that case,
the effect of random perturbations is smooth, as suggested by the 
fact that the noise parameter $\kappa$ plays the role of a friction
term in \Eq{sys1}. 

Moreover, one can see from the figures that for 
the value $E/\Delta m^2 \sim 6.7 \times 10^{4}$ eV$^{-1} \cos 2 \theta$, 
required in order for the neutrinos to undergo resonant conversion 
just at the solar centre $r = 0$, the survival probability remains equal 
to  0.5 irrespective of the $\xi$ values. The presence of this "fixed 
point" is easily understood: for such $E/\Delta m^2$ value 
\footnote{Note that the matter mixing angle is understood as the 
mixing angle that diagonalises the Hamiltonian \Eq{ev1} at each time $t$.  
It is written as $\sin^2 2\theta_{m} = 4H^2_{ex}/\omega^2_0$. At resonance 
$\omega^2_0= 4H^2_{ex}$, so that $\sin^2 2\theta_{m}=1$.} 
the neutrino state $|\nu_e> = \sqrt{2}(|\nu_{1m}> +|\nu_{2m}>)/2$ 
is produced at its resonance point and $\nu_{1m} \leftrightarrow \nu_{2m}$ 
transitions between matter eigenstates occur at the same rate. This case 
of coincidence of neutrino production point with its resonance point is 
the only one for which the effect of the matter noise is strictly absent, 
even if the adiabaticity condition holds.    

\section{Comparison with Solar Neutrino Experiments}

As seen above, there can be a substantial effect of matter 
noise fluctuations on the neutrino conversion probabilities.
It is therefore important to analyse the possible impact of this 
scenario in the determination of solar neutrino parameters
from the experimental data. The most recent averaged data of 
the chlorine \cite{cl}, gallium \cite{ga,sa} and Kamiokande 
\cite{k} experiments are:
\be[data]
R_{Cl}^{exp}= (2.55 \pm 0.25) \mbox{SNU}, \,\,\,\,
R_{Ga}^{exp}= (74 \pm 8) \mbox{SNU}, \,\,\,\
R_{Ka}^{exp}= (0.44 \pm 0.06) R_{Ka}^{BP95} 
\ee 
where  $R_{Ka}^{BP95}$ is the BP95 SSM prediction. 
For the gallium result we have taken the of GALLEX 
$R^{exp}_{Ga}= (77\pm8\pm5)$SNU\cite{ga} and SAGE 
$R^{exp}_{Ga}= (69\pm 11\pm 6)$SNU \cite{sa} measurements.
The detection rates in the chlorine and gallium experiments are
given as 
\be[rate1]
R_{Cl,Ga} = \int dE \sigma(E) \cal{P}(E)\sum_i \phi_i (E) ,
\ee
where the sum is understood over the $\nu$ source contributions
($i= ^7$Be, $^8$B...) and $\sigma(E)$ are the corresponding 
neutrino cross sections. For the Kamiokande experiment, the 
detection rate is 
\be[rate2]
R_k = \int_{Th} dE \Bigl[ \sigma_{\nu_e}(E) \cal{P}(E)
+ \sigma_{\nu_x}(E)(1- \cal{P}(E))\Bigr] \phi_B (E) ,
\ee
where $\sigma_{\nu_e}(E)$ and $\sigma_{\nu_x}(E)$ ($x=\mu,\tau$)  
are the $\nu_e\!-\!e$ and $\nu_x\!-\!e$ elastic scattering cross 
sections, respectively, and 'Th' stands for the detection energy 
threshold. In the case of sterile conversion   $\sigma_{\nu_x}=0$.    

In Fig. 2 we show the iso-signal contours (within 2$\sigma$ standard 
deviations) for each experiment and for different $\xi$ values.
These plots demonstrate that the horizontal adiabatic lines
are the ones mostly affected by the noise fluctuations. Indeed,
the larger the $\xi$ value, the greater the suppression of the 
neutrino conversion and, as a result, the larger the shift 
of this horizontal branch towards smaller $\Delta m^2$ values. 
These lower $\Delta m^2$ values allow neutrinos of lower energy 
to be involved in the adiabatic conversion (since the resonance 
matter density is proportional to $\Delta m^2/E$) so as to
compensate for the effect of the matter noise. 
Notice also that, because of this downward shift in $\Delta m^2$, 
the minimum allowed values of $\sin^2 2\theta$ for each experiment
becomes larger in order to preserve adiabaticity (see Eq. (\ref{alfa})).
Also the diagonal (so-called "non-adiabatic") and the upper
portion of the vertical (so-called large mixing) branches of the 
MSW plot are modified by the effect of noise. The diagonal lines 
are deformed mostly in the upper-left part, due to a shift of the 
kink towards larger values of the mixing angle. In contrast, they 
are less affected for $\sin^2 2\theta \gsim 10^{-2}$. This follows 
from the fact that for larger mixing the noise adiabaticity is lost 
(see Eq.(\ref{alfa})). Thus we find that the expectations derived 
on the basis of our discussion in the previous section are confirmed.

Comparing the allowed regions of all experiments shown in
Fig. 2a, 2b and 2c for the \ne $\ra$ \nm or \nt neutrino
conversion case, one concludes that the overlapping area is not 
substantially changed in the small mixing branch, whereas it 
increases in the adiabatic one,  for  large mixing 
$\sin^2 2\theta>0.3$ and $10^{-5}$eV$^2 < \Delta m^2 < 10^{-4}$eV$^2$.

We now turn to the case of $\nu_e\rightarrow\nu_s$ sterile resonant 
transitions. Here the $\nu_e$ survival probability is not substantially
changed with respect to the active neutrino conversion case, since the 
solar neutron contribution in the matter potential is rather small,
compared to that of the electrons. As a result the signal
expected in radiochemical experiments is rather insensitive to
whether the converted neutrino is active or sterile. Thus we
focus on the Kamiokande experiment. In Fig. 2d we show the iso-signal 
contours for the case of sterile neutrino conversion. One can see 
from the figure that in the sterile case, irrespective of the assumed
level of noise, the vertical large mixing branch gets thinner and 
closer to the maximal mixing region $\sin^2 2 \theta=1$. This is
required, of course, in order to increase the contribution to the 
signal which is now lost when compared to the active $\nu_{\mu,\tau}$ case.
One can also see that the noise has the same qualitative features as 
in the case of active conversions, mostly affecting the horizontal
adiabatic region of larger $\Delta m^2$.

In order to determine the solar neutrino parameters $\Delta m^2$ 
and $\sin^2 2\theta$ we now proceed to perform a $\chi^2$ analysis 
for the present experimental data.  For simplicity 
we neglect for the moment the theoretical uncertainties. 

The results of our fit are shown in Fig. 3, where the 90\% 
confidence level (C.L) areas are drawn for different $\xi$ values 
(see also Table 1). From Fig. 3a one can observe the modifications 
in the small mixing region caused by the noise in the case of active 
neutrino conversion. One sees that there is a slight shift of $\Delta m^2$ 
towards lower values and a larger shift of $\sin ^2 2 \theta$ towards larger 
values. For example the allowed region for the mixing angle covers the range
$4 \times 10^{-3}<\sin^2 2\theta<8 \times 10^{-3}$ obtained for $\xi=0$ 
becomes $8 \times 10^{-3}<\sin^2 2\theta< 2 \times 10^{-2}$ for $\xi=8\%$. 
The corresponding allowed $\Delta m^2$ range is 
$2.5 \times 10^{-6} <\Delta m^2< 9 \times 10^{-6}$ eV$^2$
to be compared with
$5 \times 10^{-6} <\Delta m^2< 1.2 \times 10^{-5}$ eV$^2$
in the noiseless case.
The large mixing area is less stable, exhibiting a tendency to shift 
towards smaller $\Delta m^2$ and $\sin^2 2 \theta$. For example, if
we take $\xi=8\%$, for the sake of argument, we find that the small 
mixing region is much more stable than the large mixing one, even 
for such a relatively large value of the noise.

As for the value of the minimal $\chi^2$, the presence of the matter 
density noise makes the data fit a little poorer: 
$\chi^2_{min}= 0.1$  for  $\xi=0$,
$\chi^2_{min}= 0.8$ for $\xi=$ 4\% and 
$\chi^2_{min}= 2.1$ for $\xi=$ 8\%. Also the best fit points 
where the $\chi^2_{min}$ is achieved change slightly: the value of 
$\Delta m^2 \sim 6 \div 7 \times 10^{-6} $eV$^2$ is almost unchanged, 
while the value of the mixing angle gets larger with respect to the
noiseless case. For example, $\sin^2 2 \theta= 6 \times 10^{-3}$ for 
$\xi=0$ while $\sin^2 2 \theta= 8 \times 10^{-3}$ for $\xi=4\%$ and
$\sin^2 2 \theta = 10^{-2}$ for $\xi=8\%$. The strong $^7$Be neutrino 
suppression, characteristic of the MSW effect, is reduced by the 
presence of matter noise (see Fig. 1). As a result, the conflict 
between chlorine and Kamiokande data is exacerbated and the data fit 
gets worse. In any case our results for $\chi^2_{min}$ (see Table 1) 
indicate that the MSW scenario still provides a good fit of the
totality of solar neutrino data, even in the presence of matter 
fluctuations, as long as $\xi \leq 8\%$ or so.

As for the large mixing solution, although the $\chi^2_{min}$ 
value is not substantially changed with respect to the noiseless case (see 
Table 1{\bf A}), we find that it acquires an increased statistical 
significance with respect to the corresponding region of the noiseless
case. Our results show that the large mixing solution gets wider
than in the noiseless case. For example the smallest allowed 
$\sin ^2 2 \theta$ value shifts from $\sin ^2 2 \theta \sim 0.4$ 
for $\xi=4\%$ down to $\sin ^2 2 \theta \sim 7 \times 10^{-2}$ for $\xi=8\%$. 

As for the best fit points we find that $\Delta m^2 \sim 10^{-5}$eV$^2$ 
is almost unchanged, whereas the best value of the mixing 
angle decreases from $\sin^2 2 \theta= 0.67$ for the noiseless case 
down to $\sin^2 2 \theta= 0.27$ for $\xi=8\%$. 
Note that the possibility of lowering the mixing angle value 
$\sin^2 2\theta$ characterising the large mixing MSW solution in 
the presence of noise may eliminate the supernova argument given in 
Ref. \cite{SSB} against such solution.
In agreement with Ref.\cite{Cala}, we find that in our fit 
this region appears already at the 80\% C.L. in the $\xi=0$ case
\footnote{This result may be underestimated since the earth 
regeneration effect has not been included.}.

We now turn to the case of sterile solar neutrino conversions. 
We find that the data fit is worse ($\chi^2_{min}= 1$) 
than for the active case (see Fig. 3b and Table 1{\bf B}) 
 and it excludes, even at 95\% C.L., the large 
mixing region (in the noiseless case). This is in agreement with 
previous analyses \cite{FIT,Cala}. However, the presence of matter 
density noise may restore this region. For example for $\xi=8\%$, 
although the data fit is much worse than in the $\xi=0$ case, the 
large mixing region appears at the 90\% C.L. We may note in this
context that the indicated range for the mixing angle is not in 
conflict with the primordial helium abundance constraints \cite{sbbn}.

So far in our analysis we have neglected SSM theoretical uncertainties, 
and worked entirely within the BP95 model \cite{SSM}. One way to account
for these uncertainties would be to allow the solar neutrino fluxes to 
vary as suggested in Ref. \cite{smirnov,BR}. 
However one can get an idea (even if partial)  of
these uncertainties by simply repeating the data fit assuming the
SSM of Turck-Chieze {\it et al.} (TCL) \cite{turck}. 
For our purposes 
the main difference between this model and the BP95 model is that it 
predicts a lower $^8$B flux. The comparison of the allowed parameter
regions obtained in the framework of the TCL model, Fig. 3 (c,d), 
with those obtained using the BP95 model, Fig. 3 (a,b), shows that 
the general features of the effect of the noise are maintained. 
In particular, our results once again establish the fact that
the indicated $\Delta m^2$ range for the small mixing MSW 
solution is fairly stable, as long as the assumed noise level
is not too large. Note also from the figures that, even though 
the effect of the noise is to lower the $\Delta m^2$ range for 
the large mixing solution, the region obtained (e.g. for $\xi = 8$\%) 
lies higher than the corresponding range for the BP95 model.


\section{Implications for Future Experiments}

Up to now we have discussed the possible consequences of the presence 
of matter fluctuations for the ongoing solar neutrino experiments.
We now turn our attention to the possibility of probing the level 
of matter noise in the Sun in the next generation of solar neutrino 
experiments.

As we have seen the $^7$Be neutrinos are the component of the
solar neutrino spectrum which is most affected by the presence 
of matter noise. Therefore the future Borexino experiment, aimed 
to detect the $^7$Be neutrino flux \cite{borex} through the elastic 
$\nu-e$ scattering should be an ideal tool for studying the solar 
matter fluctuations.

In Ref. \cite{FIT} it was shown that in the relevant (noiseless) 
MSW parameter region the Borexino signal cannot be sharply predicted.
This is illustrated in Fig. 4a, where we display the Borexino signal 
in the $\Delta m^2$ - $\sin^2 2 \theta$ plane, expressed in units of 
the expected SSM rate, i.e. $Z_{Be}=R^{pred}_{Be}/R^{SSM}_{Be}$. 
As one can see, the allowed range for the signal in this case lies 
anywhere between 0.2 to 0.7 of the SSM prediction. 
In Fig. 4b, we show the corresponding beryllium line predictions for 
the case of noisy MSW, assuming $\xi= 4\%$. We see that the presence 
of matter noise strongly modifies the picture: the minimal allowed 
value for $Z_{Be}$ now becomes higher, $Z_{Be}\geq 0.37$. Therefore 
if the Borexino experiment detects a small signal, $Z_{Be} \lsim 0.3$
(with sufficient accuracy) this will imply that a 4\% level of matter 
fluctuations in the central region of the Sun is rather unlikely to be 
present if the MSW mechanism is responsible for the explanation of the
solar neutrino deficit
\footnote{In principle any value of $Z_{Be}$ is also compatible 
with the just-so oscillation scenario \cite{BR,Cala}, but here the 
strong seasonal $^7$Be and $pep$ signal variations, would help
to distinguish from the MSW case.}. 

Note, on the other hand, that if a higher value $Z_{Be} \gsim 0.5$
would be found experimentally, this would be incompatible with the 
small mixing MSW solution with noise at the $\xi = 4\%$ level.
However, this higher signal could be consistent with the both
the large mixing MSW solution as well as the noiseless small 
angle MSW solution. On the other hand, if the noise level is 
higher, $\xi = 8\%$, the allowed $Z_{Be}$ range narrows down 
to values between 0.5 to 0.65. 

Let us turn to the case sterile resonant conversion in the noisy 
MSW effect. Let us imagine that future large detectors such as 
Super-Kamiokande and/or the Sudbury Neutrino Observatory (SNO) 
establish through, e.g. the measurement of the charged to neutral 
current ratio, that the deficit of solar neutrinos is due to the 
\ne $\ra$ \ns resonant conversion. In this case, the minimum signal 
expected in Borexino is very small $Z_{Be} \approx 0.02$ for 
$\xi =0$ (see Fig. 4c). 
On the other hand in the noisy case with $\xi=4\%$, 
the minimum expected Borexino signal is 10 times higher than in the
noiseless case, so that if Borexino detects a rate $Z_{Be} \lsim 0.1$ 
(see Fig. 4d) this would again exclude noise levels above  $4\%$.

\section{Discussion and Conclusions}

We have presented a comprehensive study of the effects of the matter 
density noise upon the MSW solution to the solar neutrino problem. 
We have adopted the wave function Schr\"odinger formalism to re-write 
the corresponding MSW evolution equations for the neutrino survival 
probabilities. The fluctuations weaken the efficiency of the MSW 
suppression in the adiabatic regime, whereas they are much less
effective in the non-adiabatic regime. In our data fit we have shown 
that the MSW solution still exists for realistic levels of matter 
density noise $\xi \lsim 8\%$. However, our $\chi^2$ analysis has 
shown that the quality of the fit gets a little worse if these
noisy matter perturbations are present. In any case the mass range 
determined from our fit for the small mixing MSW solution 
$4 \times 10^{-6} \mbox{eV}^2 <\Delta m^2< 10^{-5}\mbox{eV}^2$ 
is relatively stable at 90\% C.L., whereas the mixing angle determination 
appears more sensitive to the assumed level of fluctuations, and
shifts $\sin^2 2 \theta$ towards larger values up to  $10^{-2}$. 
These trends also hold for the case of sterile solar neutrino conversion. 
In the latter case we have found that in the presence of solar density 
noise the large mixing region gets somewhat improved statistical 
significance when compared with the noiseless case. However, 
it is remains highly disfavoured with respect to the small
mixing MSW solution.

We have also explored the potential of the Borexino experiment to "test" 
the level of matter density fluctuations in the solar interior through the 
measurement of the $^7$Be neutrino flux, as depicted in Fig. 4.

Finally, we note that in our analysis we have neglected
the details of the neutrino production distribution as a function
of the distance to the solar centre. It is well known that this affects 
mainly the low energy $pp$ neutrinos \cite{MSW}. As a result, the 
iso-signal curves we have obtained for the gallium experiments are 
somewhat less reliable in the position of the kink corresponding to 
$\Delta m^2 \lsim 2 \times 10^{-6} $eV$^2$ marking the onset of
$pp$ neutrino suppression and lying on the gallium non-adiabatic 
branch. However, this does not substantially affect the 
determination of the relevant regions where {\sl all} solar neutrino 
data are explained through the MSW effect. This includes both
small as well as large mixing MSW solutions.

\vskip 1truecm

\section*{Acknowledgements}
We thank Z. Berezhiani, N. Hata P. Krastev, S. Mikheyev and A. Smirnov for 
valuable comments and discussion. We also thank S. Turck-Chi$\acute{e}$ze 
for an informative conversation. This work has been supported by DGICYT 
under Grant numbers PB92-0084, SAB94-0325 and by RFFR-95-02-03724 (V. S.),
by the grant N. ERBCHBI CT-941592 of the Human Capital and Mobility 
Program (A. R.) and by a DGICYT postdoctoral fellowship (H. N.)

\newpage


\vspace{-0.5cm}
{\small 
\begin{table}[hbt]
\begin{center}

\begin{tabular}{|l|c|c|c|c|} \hline 
~{\large\bf A:} ${\it active}$  & $\xi=0 $ & 
$\xi =2\%$ & $\xi=4\%$ & $\xi=8\%$  \\ 
\cline{1-5}
$\mbox{small}\,\theta$ & & & &\\
\hline
$\chi^2_{min}$ & 0.10 & 0.23 & 0.80 & 2.1 \\ 
\hline
$\Delta m^2(10^{-5}\mbox{eV}^2)$ & 0.68 & 0.71 & 0.65 & 0.61 \\
\hline
$\sin^2 2 \theta$ & $6.2 \times 10^{-3}$ & $7.3 \times 10^{-3}$ & 
$7.5 \times 10^{-3}$ & $10^{-2}$ \\
\cline{1-5}
\cline{1-5}
$\mbox{large}\,\theta$ & & & &\\
\hline
$\chi^2_{min}$ & 3.3 & 2.9 & 3.0 & 3.2 \\ 
\hline
$\Delta m^2(10^{-5}\mbox{eV}^2)$ & 2.7 & 2.4 & 2.0 & 1.2 \\
\hline
$\sin^2 2 \theta$ & $0.67$ & $0.69$ & 
$0.57$ & $0.27$ \\
\hline
\end{tabular}
\label{tab1a}
\end{center}
\end{table}
     
\vspace{-1.0cm}

\begin{table}[hbt]
\begin{center}
\begin{tabular}{|l|c|c|c|c|} \hline 
~{\large\bf B:} ${\it sterile}$  & $\xi=0 $ & $\xi=2\% $ & $\xi=4\%$ &
$\xi=8\% $   \\ 
\cline{1-5}
$\mbox{small}\,\theta$  & & & &\\
\hline
$\chi^2_{min}$ & 1.0 & 1.9 & 3.6 & 8.9  \\ 
\hline
$\Delta m^2(10^{-5}\mbox{eV}^2)$ & 0.53 & 0.50& 0.49 & 0.40  \\
\hline
$\sin^2 2 \theta$ & $7.5 \times 10^{-3}$ & $7.5 \times 10^{-3} $ &
 $9.0 \times 10^{-3}$ & $1.3\times 10^{-2}$  \\
\cline{1-5}
\cline{1-5}
$\mbox{large}\,\theta$ & & & & \\
\hline
$\chi^2_{min}$ & 10 & 11 & 11 & 11  \\ 
\hline
$\Delta m^2(10^{-5}\mbox{eV}^2)$ & 1.4 & 1.2 & 1.6 & 1.0 \\
\hline
$\sin^2 2 \theta$ & $0.83$ & $0.83$ & 0.69& 0.39  \\
\hline
\end{tabular}
\label{tab1b}
\end{center}
\end{table}

\vskip 1.0truecm

{\bf Table 1.} The values of $\chi^2_{min}$, for 1 degree of freedom,  
and the corresponding best fit $\Delta m^2$ and $\sin^2 2 \theta$ 
parameters in the small and large mixing regions, for different 
values of $\xi$. 
Tables {\bf A} and {\bf B} are for the active-active and active-sterile 
conversion respectively, using the latest 1995 Bahcall-Pinsonneault (BP95)
model.

\newpage

\vspace{-0.5cm}
{\small 
\begin{table}[hbt]
\begin{center}

\begin{tabular}{|l|c|c|c|c|} \hline 
~{\large\bf A:} ${\it active}$  & $\xi=0 $ & 
$\xi =2\%$ & $\xi=4\%$ & $\xi=8\%$  \\ 
\cline{1-5}
$\mbox{small}\,\theta$ & & & &\\
\hline
$\chi^2_{min}$ & 0.10 & 0.46 & 1.1 & 3.0 \\ 
\hline
$\Delta m^2(10^{-5}\mbox{eV}^2)$ & 0.64 & 0.66 & 0.62 & 0.66 \\
\hline
$\sin^2 2 \theta$ & $3.6 \times 10^{-3}$ & $3.6 \times 10^{-3}$ & 
$4.3 \times 10^{-3}$ & $5.2 \times 10^{-3}$ \\
\cline{1-5}
\cline{1-5}
$\mbox{large}\,\theta$ & & & &\\
\hline
$\chi^2_{min}$ & 5.2 & 5.2 & 5.0 & 4.3 \\ 
\hline
$\Delta m^2(10^{-5}\mbox{eV}^2)$ & 17 & 16 & 12 & 3.8 \\
\hline
$\sin^2 2 \theta$ & $0.83$ & $0.83$ & 
$0.83$ & $0.69$ \\
\hline
\end{tabular}
\label{tab2a}
\end{center}
\end{table}
     
\vspace{-1.0cm}

\begin{table}[hbt]
\begin{center}
\begin{tabular}{|l|c|c|c|c|} \hline 
~{\large\bf B:} ${\it sterile}$  & $\xi=0 $ & $\xi=2\% $ & $\xi=4\%$ &
$\xi=8\% $   \\ 
\cline{1-5}
$\mbox{small}\,\theta$  & & & &\\
\hline
$\chi^2_{min}$ & 0.69 & 2.1 & 2.3 & 5.6  \\ 
\hline
$\Delta m^2(10^{-5}\mbox{eV}^2)$ & 0.51 & 0.45& 0.48 & 0.57  \\
\hline
$\sin^2 2 \theta$ & $4.3 \times 10^{-3}$ & $4.2 \times 10^{-3} $ &
 $5.2 \times 10^{-3}$ & $6.3 \times 10^{-3}$  \\
\cline{1-5}
\cline{1-5}
$\mbox{large}\,\theta$ & & & & \\
\hline
$\chi^2_{min}$ & 9.3 & 9.3 & 9.2 & 8.3  \\ 
\hline
$\Delta m^2(10^{-5}\mbox{eV}^2)$ & 16 & 15 & 14 & 3.3 \\
\hline
$\sin^2 2 \theta$ & $0.83$ & $0.83$ & 0.83& 0.69  \\
\hline
\end{tabular}
\label{tab2b}
\end{center}
\end{table}

\vglue 1cm

{\bf Table 2.} 
The same as for Table 1, but using the Turck-Chieze {\it et al.} (TCL) SSM. 

\newpage

{\bf Figure Captions}
\vskip 0.5truecm
{\bf Fig. 1}. \\
Averaged solar neutrino survival probability $\cal{P}$ versus 
$E/\Delta m^2$ for small mixing (a: $\sin^2 2 \theta= 10^{-2}$) 
and large mixing (b: $\sin^2 2 \theta= 0.7$). The solid, dotted,
dashed and dot-dashed curves correspond to noise levels 
$\xi=0, 2\%, 4\%$ and $8\%$, respectively. 
\vskip 0.5truecm
{\bf Fig. 2. }\\
Iso-rate contours for the chlorine (a), gallium (b) and $\nu-e$ 
scattering (c) experiments for the case of active neutrino conversion, 
$\nu_e \rightarrow \nu_{\mu,\tau}$. The threshold energy for the recoil 
electron detection is 7.5 MeV. For the radiochemical experiments the 
results are in SNU, whereas for the  $\nu-e$ scattering experiment 
these are given in units of the BP95 SSM prediction. The contours 
delimit the 2$\sigma$ allowed regions. The solid, dashed and dotted 
curves correspond to $\xi=0, 4\%$ and $8\%$, respectively. Fig. 2d gives 
the same iso-rate contours for the case of sterile $\nu_e \ra \nu_s$ 
conversion for the $\nu-e$ scattering experiment. 
\vskip 0.5truecm
{\bf Fig. 3. }\\
The 90\% C.L. allowed regions (given by the condition $\chi^2 \leq 
\chi^2_{min} + 4.61$) for the active  (a and c) conversion and 
for the sterile (b and d) conversion. For Fig. 3a and 3b, the C.L. 
allowed regions are obtained using the most recent Bahcall and 
Pinsonneault (BP95) SSM; for Fig. 3c and 3d we used the Turck-Chieze 
and Lopes (TCL) SSM. In Fig. 3a,  3b, 3c and 3d the solid, dot, dash, 
dot-dash curves correspond to the cases $\xi=0, 2\%, 4\%$ and $8\%$, 
respectively. The $\chi^2_{min}$ and the corresponding ($\Delta m^2$, 
$\sin^2 2\theta$) best fit points are given in Table 1 and Table 2. 
\vskip 0.5truecm
{\bf Fig. 4. }\\
The iso-signal contours of the ratio $Z_{Be}=R^{pred}_{Be}/R^{SSM}_{Be}$ 
(figures at the curves) 
in the $\nu-e$ scattering Borexino detector (solid lines).  
The threshold energy for the recoil electron detection is 0.25 MeV. 
The 90\% C.L. allowed regions (dotted lines) and 
the corresponding best fit points (diamonds)  
are also superimposed, 
as determined by the present experimental data and using the BP95 SSM. 

The case of active resonant conversion is presented in Fig. 4a and 
Fig. 4b for $\xi=0$ and $\xi=4\%$, respectively. Analogously, Fig. 4c 
($\xi=0$) and Fig. 4d ($\xi=4\%$) refer to the 
case of sterile neutrino resonant conversion.

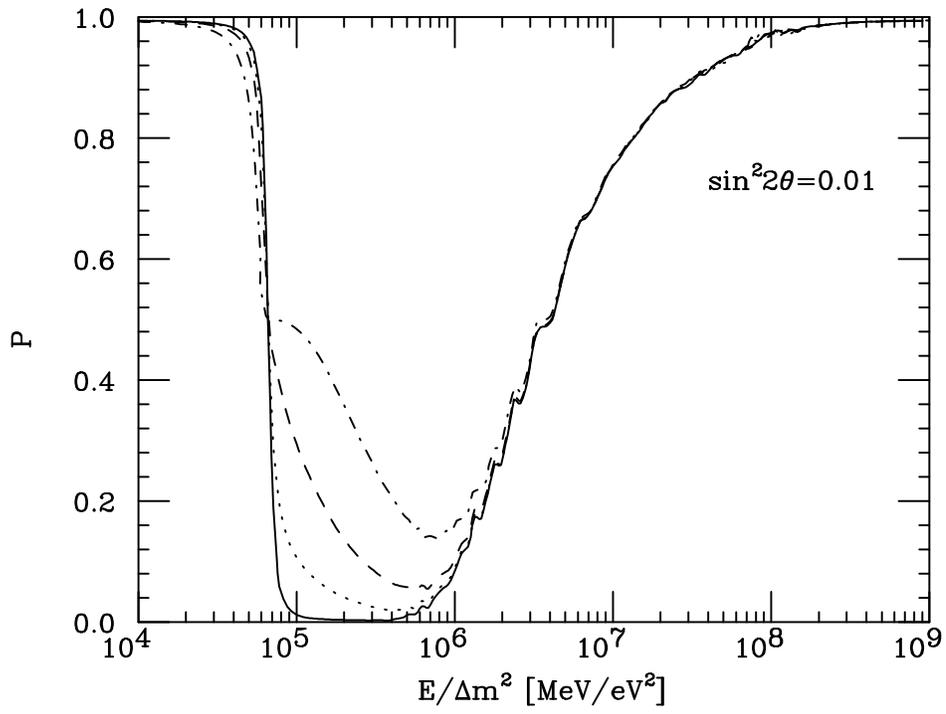
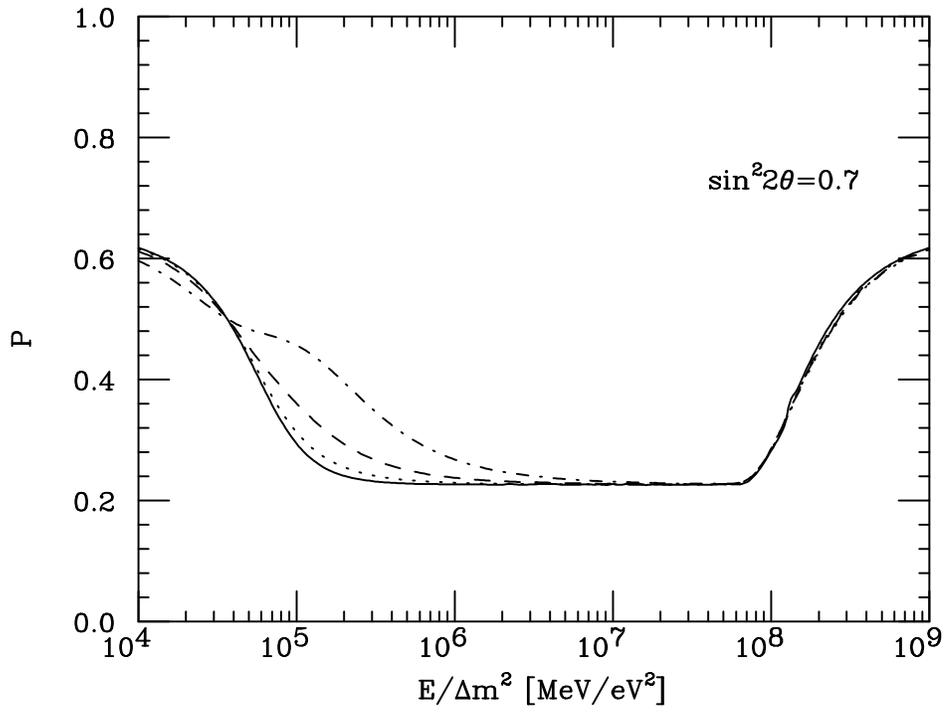
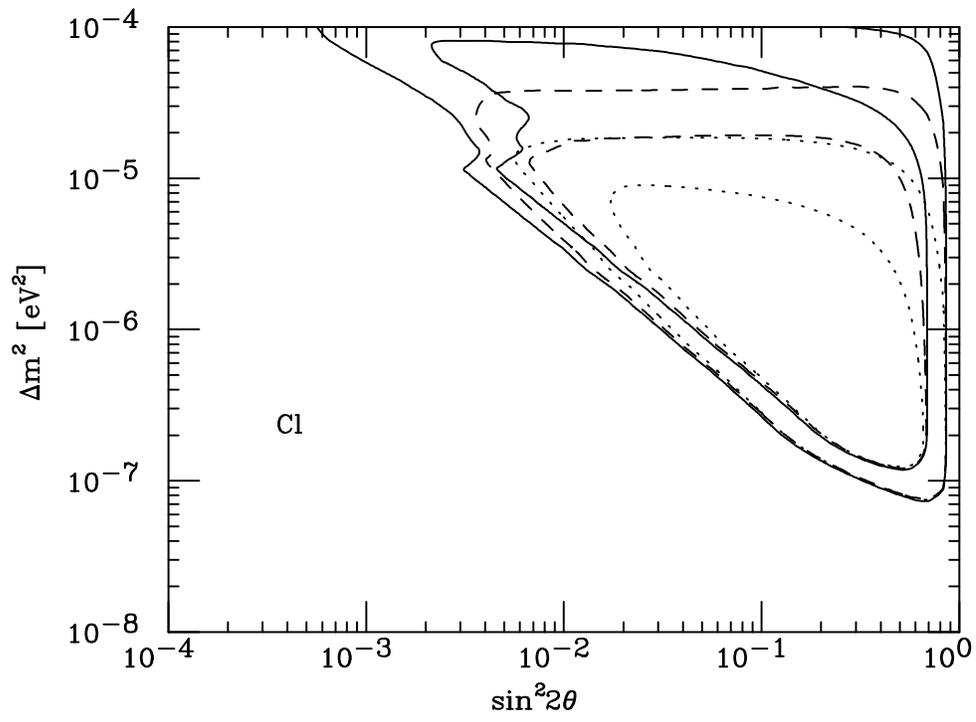
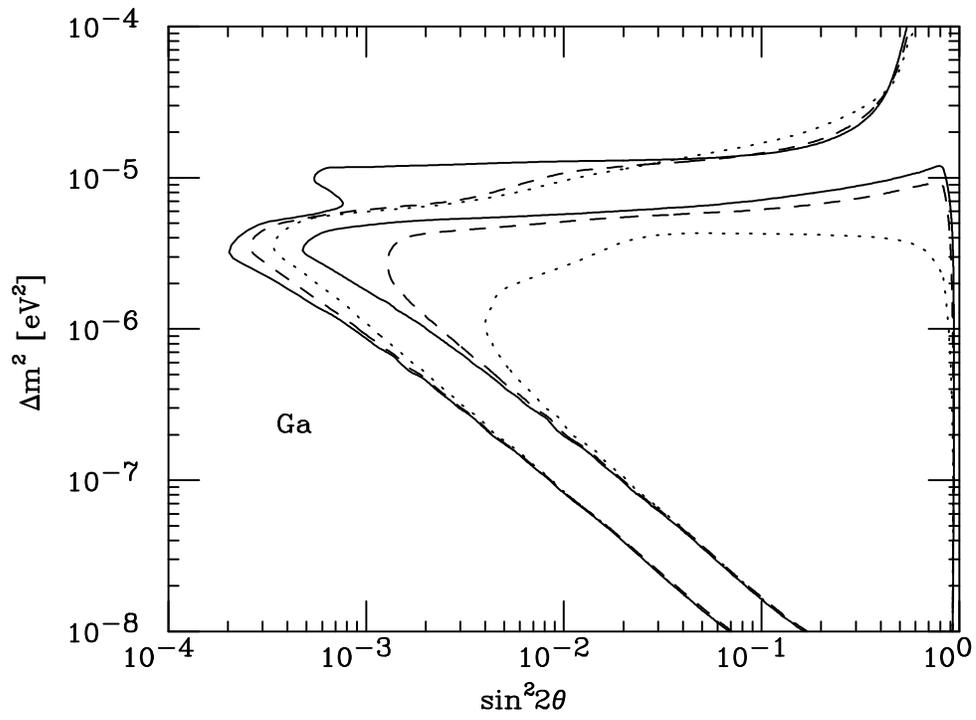
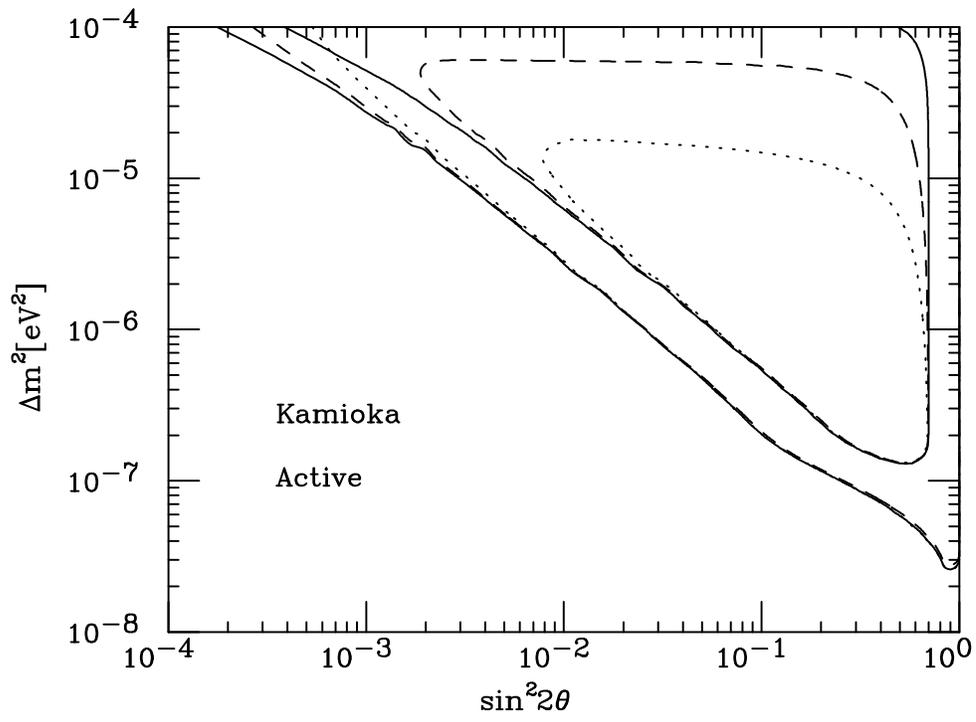
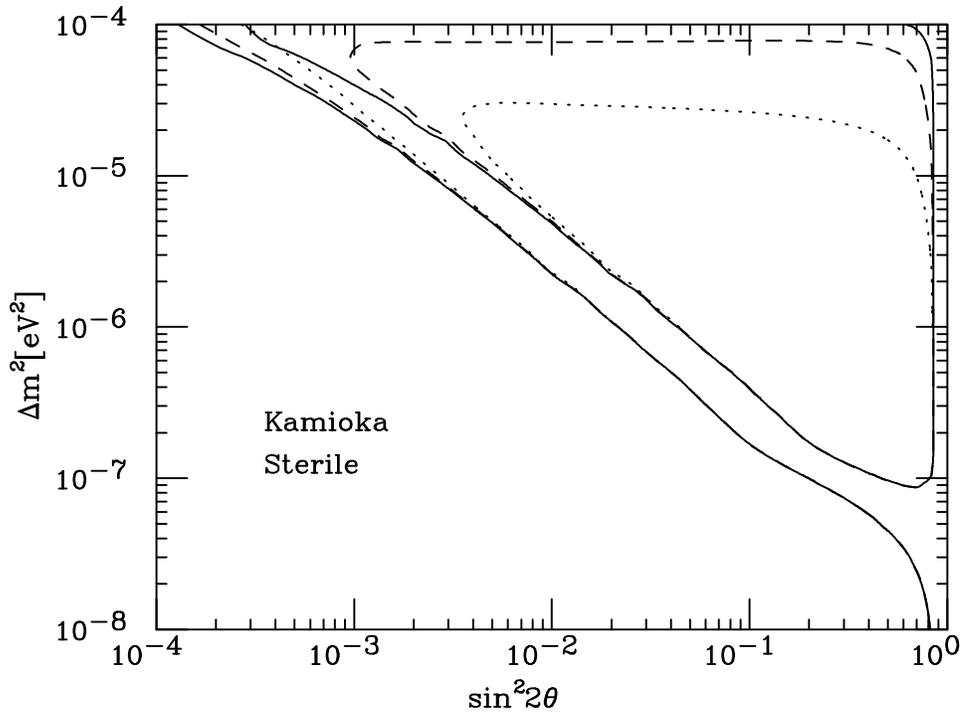
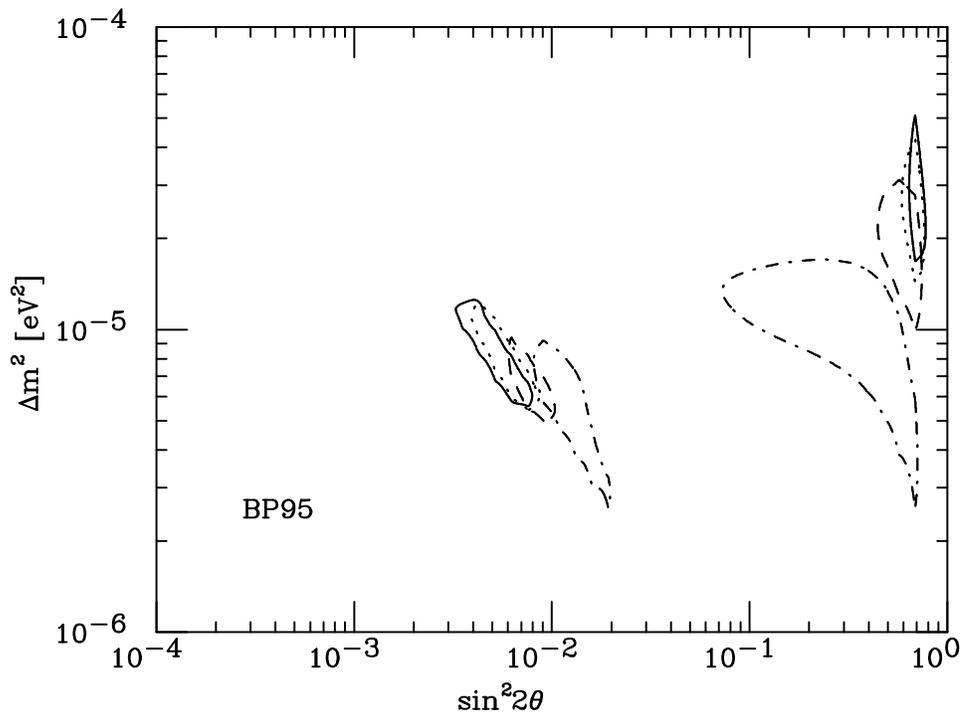
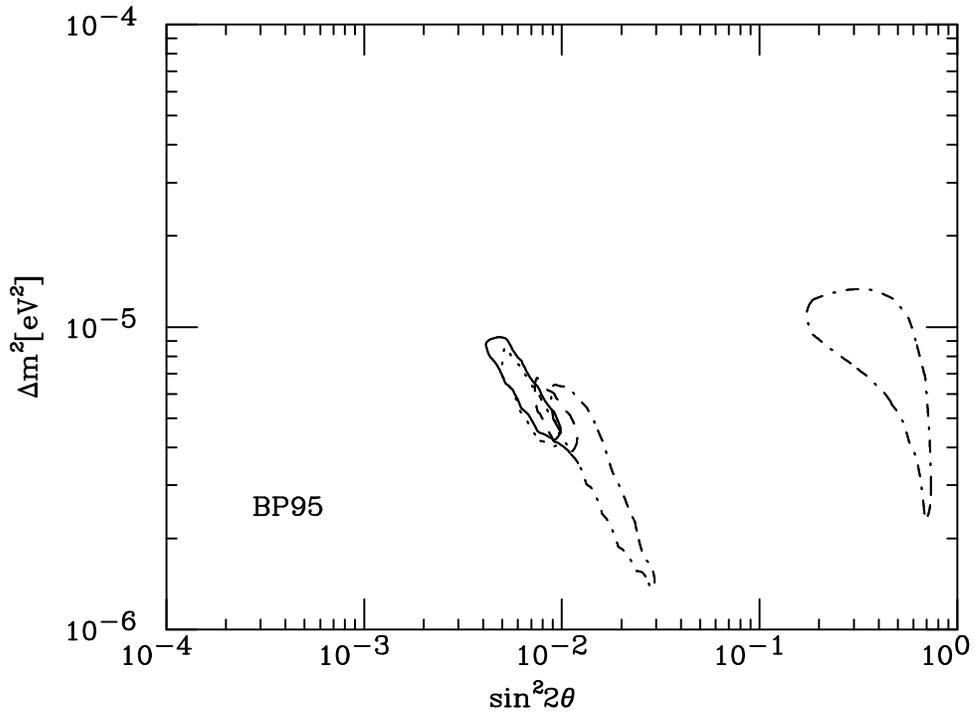
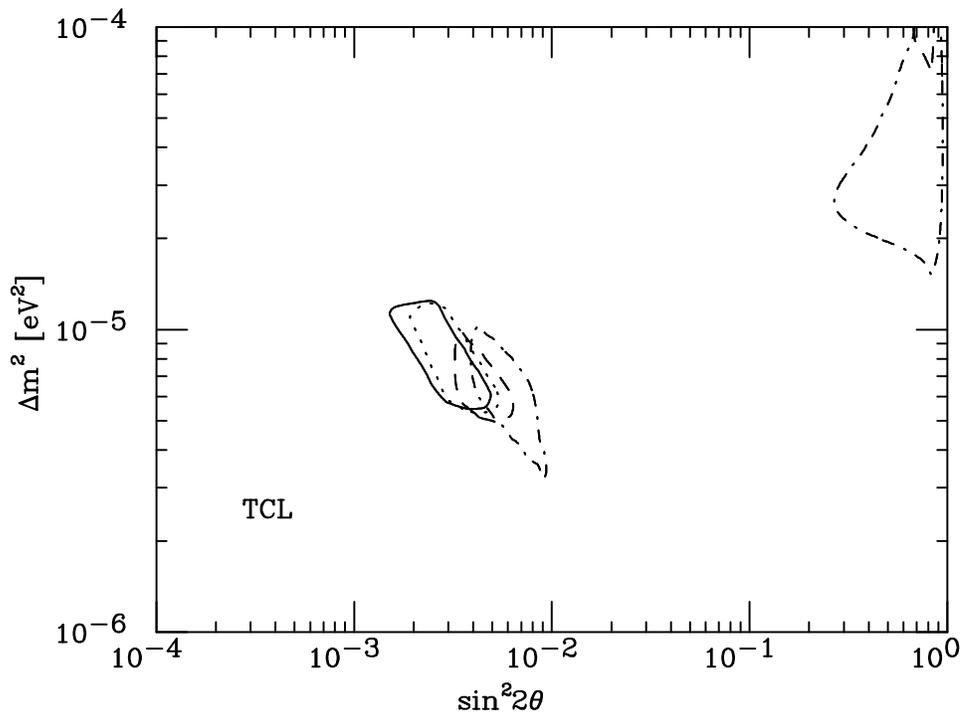
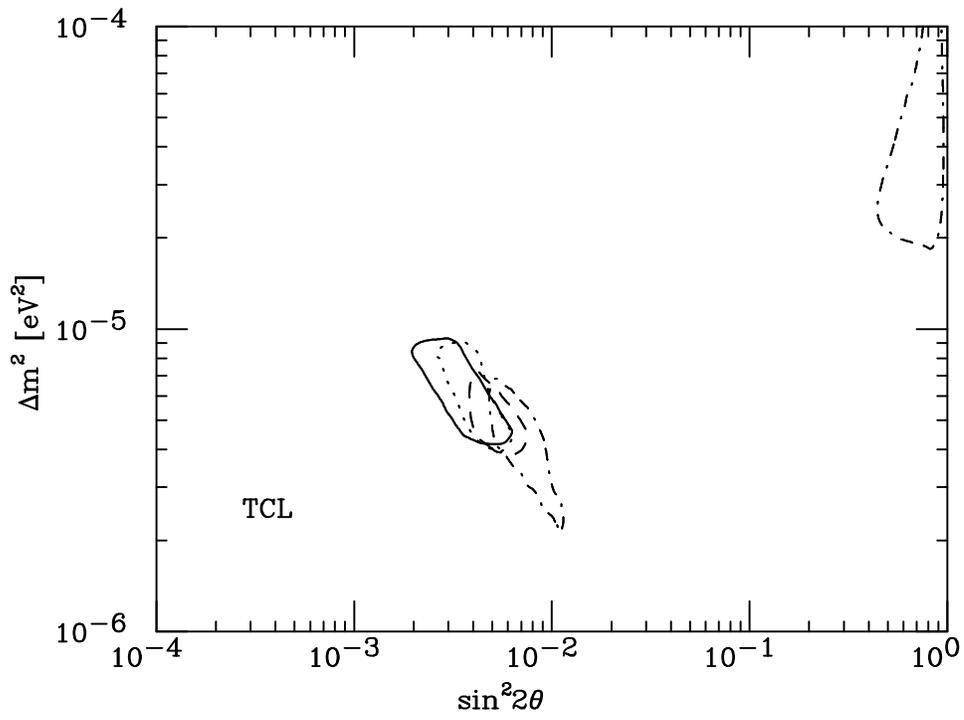
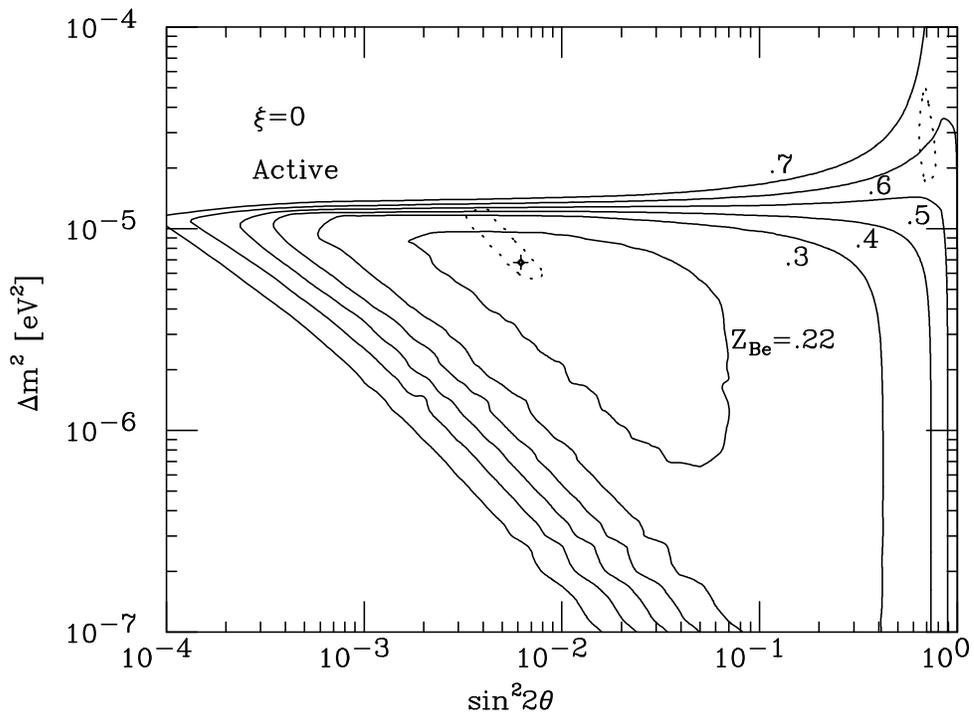
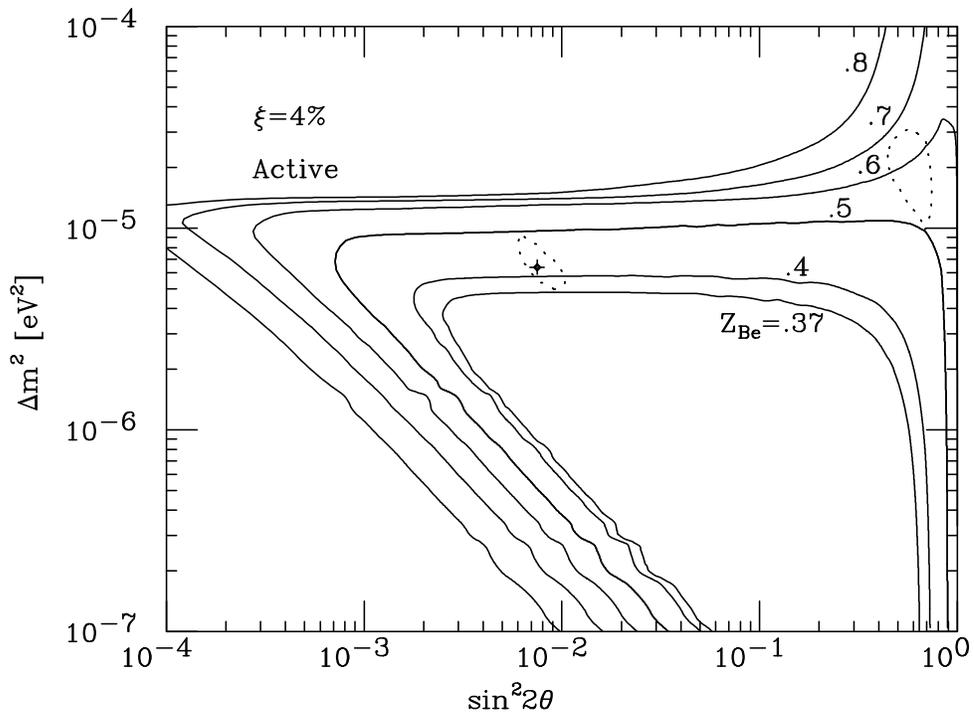
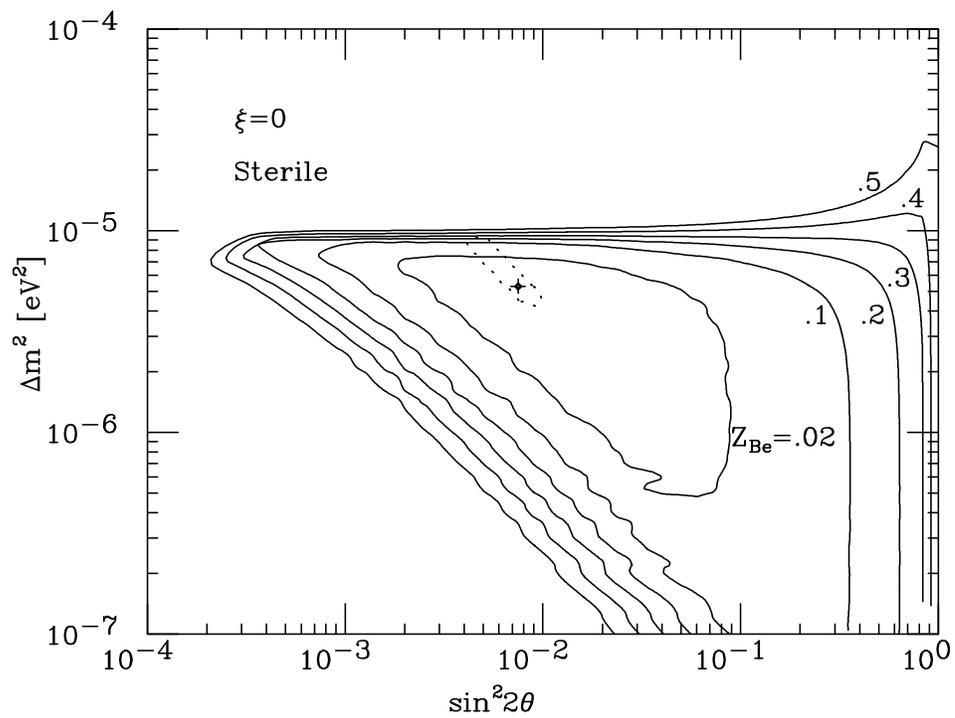
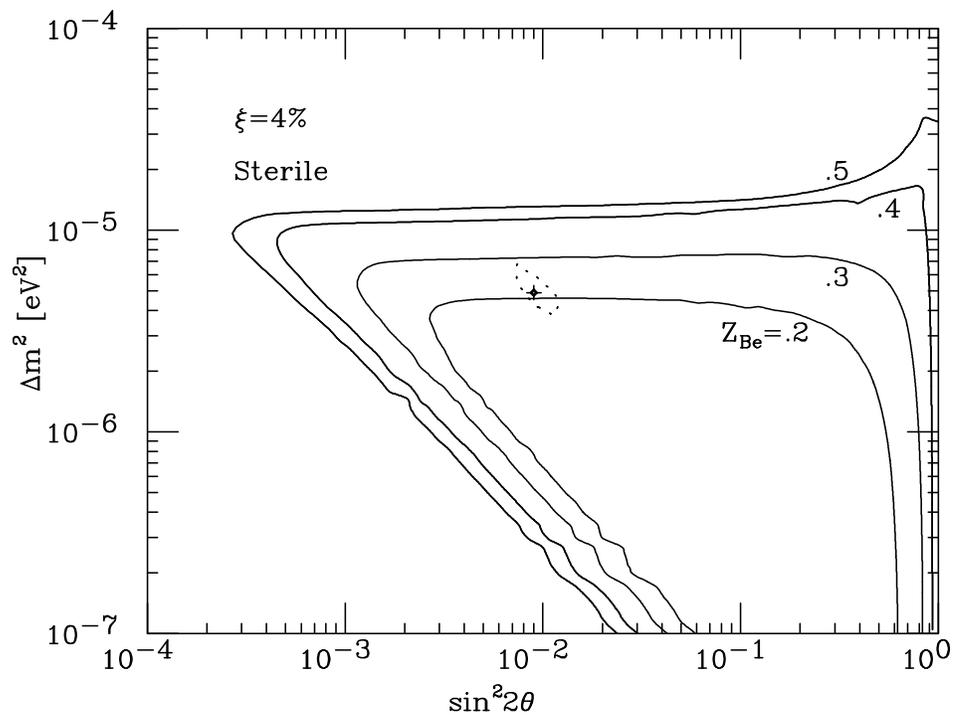
\end{document}